\RequirePackage{rotating}
\documentclass[10pt, conference, compsocconf]{IEEEtran}
\usepackage{graphicx,mathtools,multirow,cite}
\usepackage{booktabs}
\usepackage{amssymb}
\usepackage{placeins}
\usepackage{rotating}
\usepackage{afterpage}
\usepackage{fixltx2e}

\ifCLASSINFOpdf
\else
\fi
\begin{document}
%
\title{The Best of Both Worlds: Using Automatic Detection and Limited Human Supervision to Create a Homogenous Magnetic Catalog Spanning Four Solar Cycles}




%
\author{\IEEEauthorblockN{A.\ Mu\~noz-Jaramillo\IEEEauthorrefmark{1},
Z.\ A.\ Werginz\IEEEauthorrefmark{1}\IEEEauthorrefmark{2},
J.\ P.\ Vargas-Acosta\IEEEauthorrefmark{3}\IEEEauthorrefmark{2},
M.\ D.\ DeLuca\IEEEauthorrefmark{4}\IEEEauthorrefmark{2},
J.\ C.\ Windmueller\IEEEauthorrefmark{2},
J.\ Zhang\IEEEauthorrefmark{5},\\
D.\ W.\ Longcope\IEEEauthorrefmark{2},
D.\ A.\ Lamb\IEEEauthorrefmark{6},
C.\ E.\ DeForest\IEEEauthorrefmark{6},
S.\ Vargas-Dominguez\IEEEauthorrefmark{3},
J.\ W.\ Harvey\IEEEauthorrefmark{7}, \&
P.\ C.\ H.\ Martens\IEEEauthorrefmark{1}\\}
\IEEEauthorblockA{\IEEEauthorrefmark{1}Georgia State University, Atlanta, GA 30303, USA. Email: amunozj@gsu.edu}
\IEEEauthorblockA{\IEEEauthorrefmark{2}Montana State University, Bozeman, MT 95717, USA.}
\IEEEauthorblockA{\IEEEauthorrefmark{3}Universidad Nacional de Colombia, Bogot\'a, DC 111321, Colombia.}
\IEEEauthorblockA{\IEEEauthorrefmark{4}University of Colorado, Boulder, CO 80309, USA.}
\IEEEauthorblockA{\IEEEauthorrefmark{5}George Mason University, Fairfax, VA 22030, USA.}
\IEEEauthorblockA{\IEEEauthorrefmark{6}Southwest Research Institute, Boulder, CO 80302, USA.}
\IEEEauthorblockA{\IEEEauthorrefmark{7}National Solar Observatory, Boulder, CO 80303, USA.}
}


\maketitle

\begin{abstract}
Bipolar magnetic regions (BMRs) are the cornerstone of solar variability.  They are tracers of the large-scale magnetic processes that give rise to the solar cycle, shapers of the solar corona, building blocks of the large-scale solar magnetic field, and significant contributors to the free-energetic budget that gives rise to flares and coronal mass ejections.  Surprisingly, no homogeneous catalog of BMRs exists today, in spite of the existence of systematic measurements of the magnetic field since the early 1970's. The purpose of this work is to address this deficiency by creating a homogenous catalog of BMRs from the 1970's until the present.

For this purpose, in this paper we discuss the strengths and weaknesses of the automatic and manual detection of BMRs and how both methods can be combined to form the basis of our Bipolar Active Region Detection (BARD) code and its supporting human supervision module.  At present, the BARD catalog contains more than 10,000 unique BMRs tracked and characterized during every day of their observation.  Here we also discuss our future plans for the creation of an extended multi-scale magnetic catalog combining the SWAMIS and BARD catalogs.
\end{abstract}


%
\IEEEpeerreviewmaketitle

\section{Introduction: Bipolar Magnetic Regions vs.\ Active Regions}

One of the most prominent characteristics of the solar cycle is the emergence of bipolar patches of strong magnetic fields (often with a visible signature in the form of sunspots).  These bipolar magnetic regions (BMRs) have an important, but subtle duality that changes the way they are studied depending on the application.  On the one hand, the fact that they have large amounts of free energy that is eventually released in the form flares and coronal mass ejections (CMEs) makes them one of the cornerstones of space weather studies.  On the other hand, there is now strong evidence that their emergence and decay during the course of a given solar cycle plays a critical role in building the large-scale magnetic field that seeds the subsequent cycle\cite{munoz-etal2013a,cameron-schussler2015}.


\begin{figure*}[!t]
\centering
\includegraphics[height=0.32\textwidth]{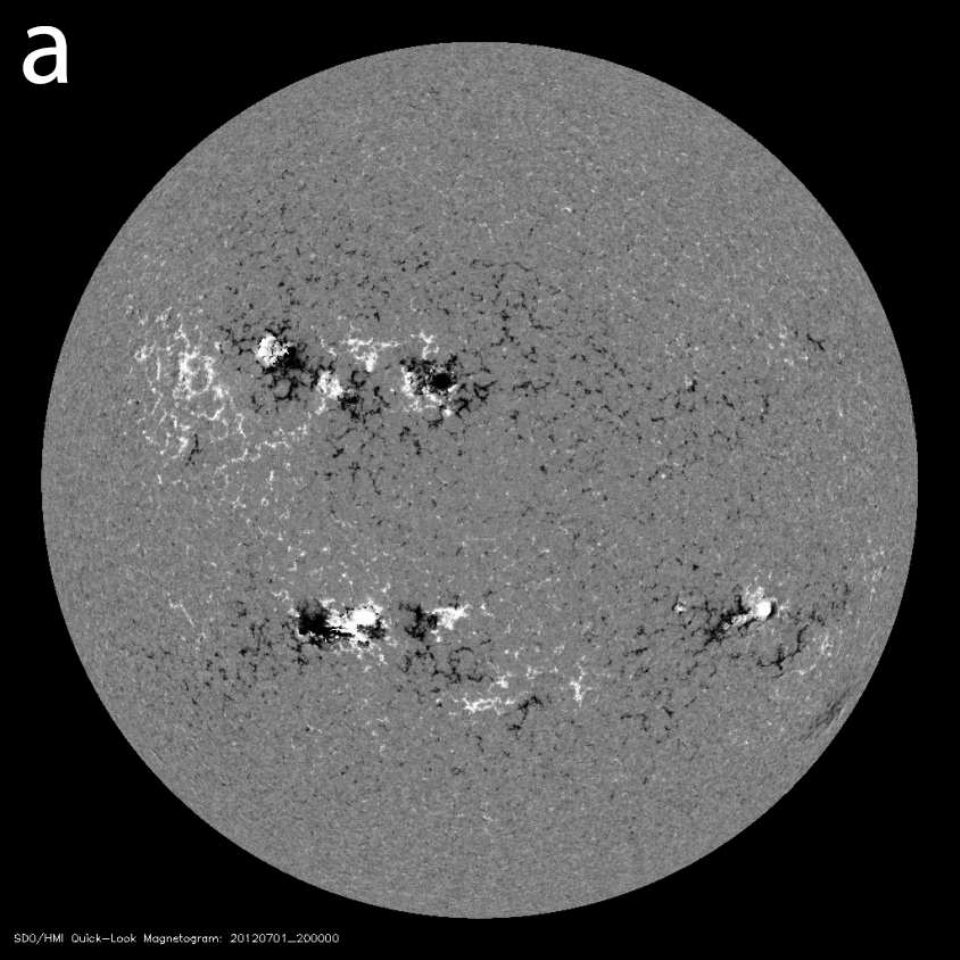}
\includegraphics[height=0.32\textwidth]{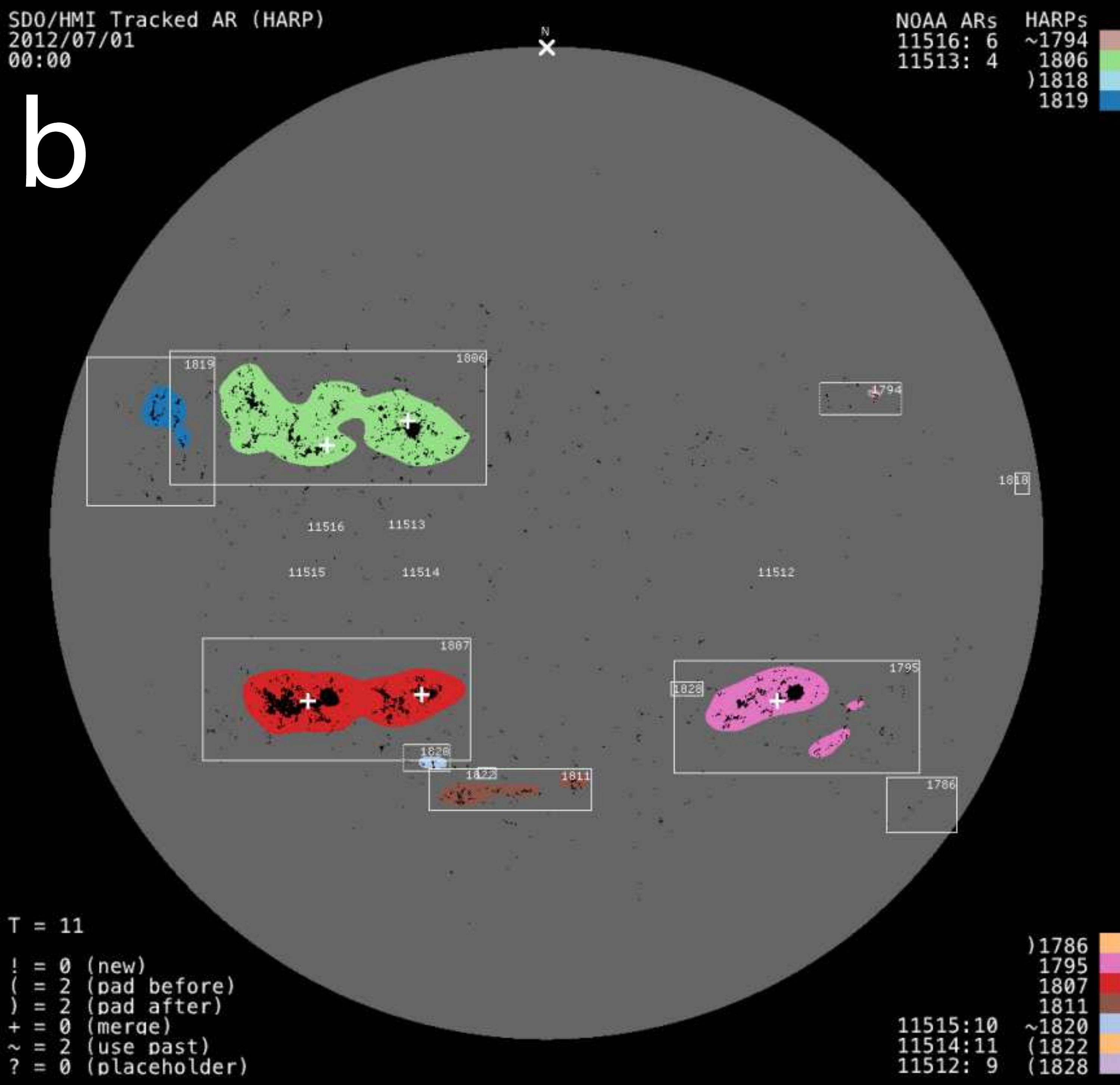}
\includegraphics[height=0.32\textwidth]{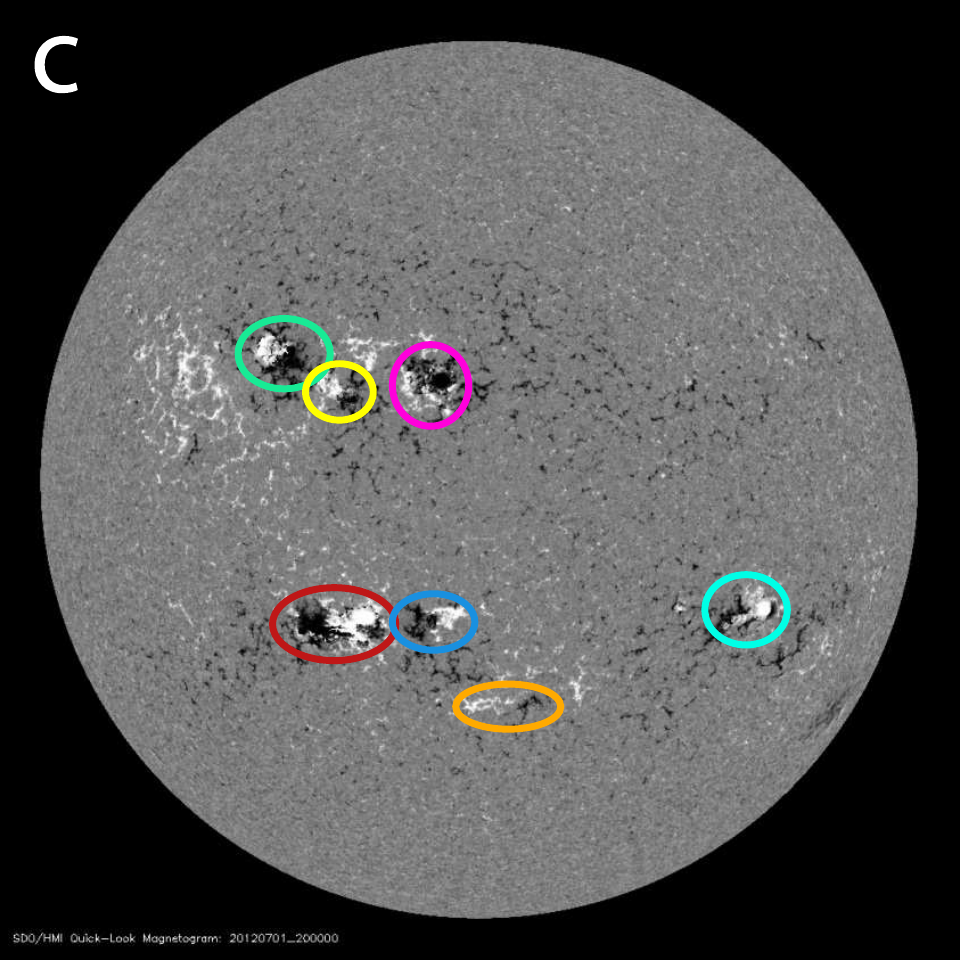}
\caption{(a) SDO/HMI magnetogram taken on 2012-Jul-07. (b) Automatic detection and masking of active regions by the HARP algorithm.  (c) Manual circling of bipolar magnetic regions.}
\label{fig_BMR_AR}
\end{figure*}

Due to the fact that these magnetic patches are directly related with the vast releases of energy that drive heliospheric variability, people often refer to them as \emph{active regions (ARs)}.  Unfortunately, this has led to a loosening of terminology because there is a subtle but important difference between BMRs and ARs:  \emph{BMRs denote the bipolar patches that appear on the surface of the Sun after a flux-emergence event.  ARs denote active complexes from which space weather events originate, which often include more than one BMR.}  This difference becomes quite important for designing automatic detection algorithms because their success relies heavily on a clear definition of the nature of the detection targets.

The difference between ARs and BMRs is nicely illustrated in Figure \ref{fig_BMR_AR}, which shows an SDO/HMI magnetogram (Fig.~\ref{fig_BMR_AR}-a), the output of applying the HMI Active Region Patches (HARP) automatic detection algorithm\cite{hoeksema-etal2014} to that magnetogram (Fig.~\ref{fig_BMR_AR}-b), and the same magnetogram showing individual BMRs circled by us by hand (Fig.~\ref{fig_BMR_AR}-c):  even though there is one perfect match between a HARP and a BMR, two of the HARPs contain more than one BMR and one HARP contains a predominantly unipolar patch.

To this date, and to the extent of our knowledge, the great majority of automatic detection algorithms used in solar physics are optimized for the detection of ARs (i.e.\ they are mainly designed with a space weather application in mind; see \cite{attie-innes2015} for an exception).  Additionally, in spite of the availability of systematic measurements of the magnetic field since the early 1970's, there is no homogenous (i.e.\ cross-calibrated catalog spanning 40 years of observations) BMR or AR catalog extending that far.  Given that our main interest is studying the role of BMRs as one of the critical building blocks of the solar cycle, the main goal of this paper is to showcase the results of our attempts to address this deficiency.

The core of our detection method is the Bipolar Active Region Detection (BARD) code, which we introduce in Section \ref{sec_BARD}.  In Section \ref{sec_mam_mach} we discuss the strengths and weaknesses of purely automatic and purely manual methods of BMR detection, as well as how to interface a human supervision module with BARD in order to maximize the accuracy of our catalog.  Our preliminary results are showcased in Section \ref{sec_results} and a discussion of our future plans to combine the BARD and SWAMIS catalogs is included in Section \ref{sec_SWAMIS}.  We finish with our summary and concluding remarks in Section \ref{sec_conclusions}.

\begin{figure*}[!t]
\centering
\begin{tabular}{ccc}
\includegraphics[width=0.28\textwidth]{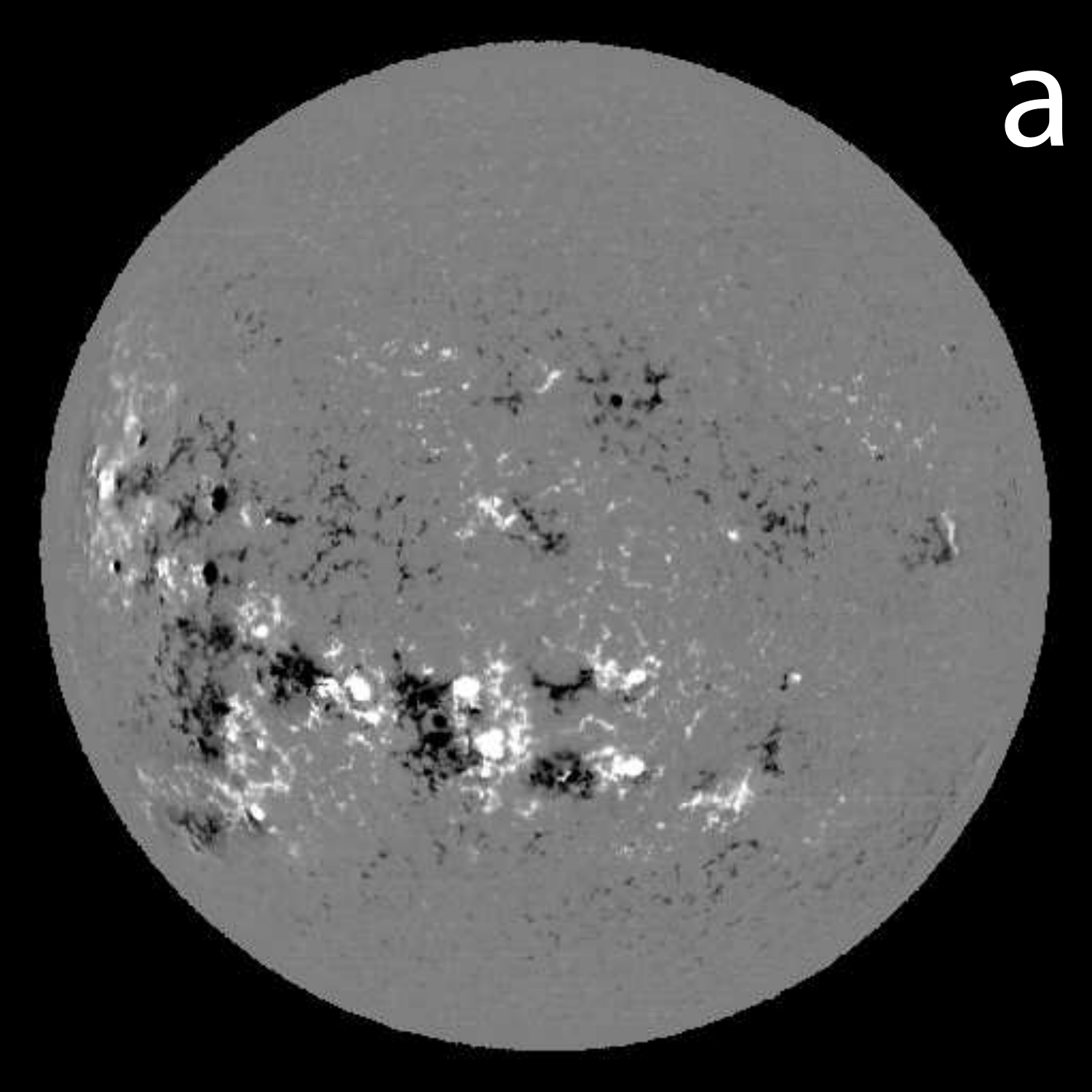}&
\includegraphics[width=0.28\textwidth]{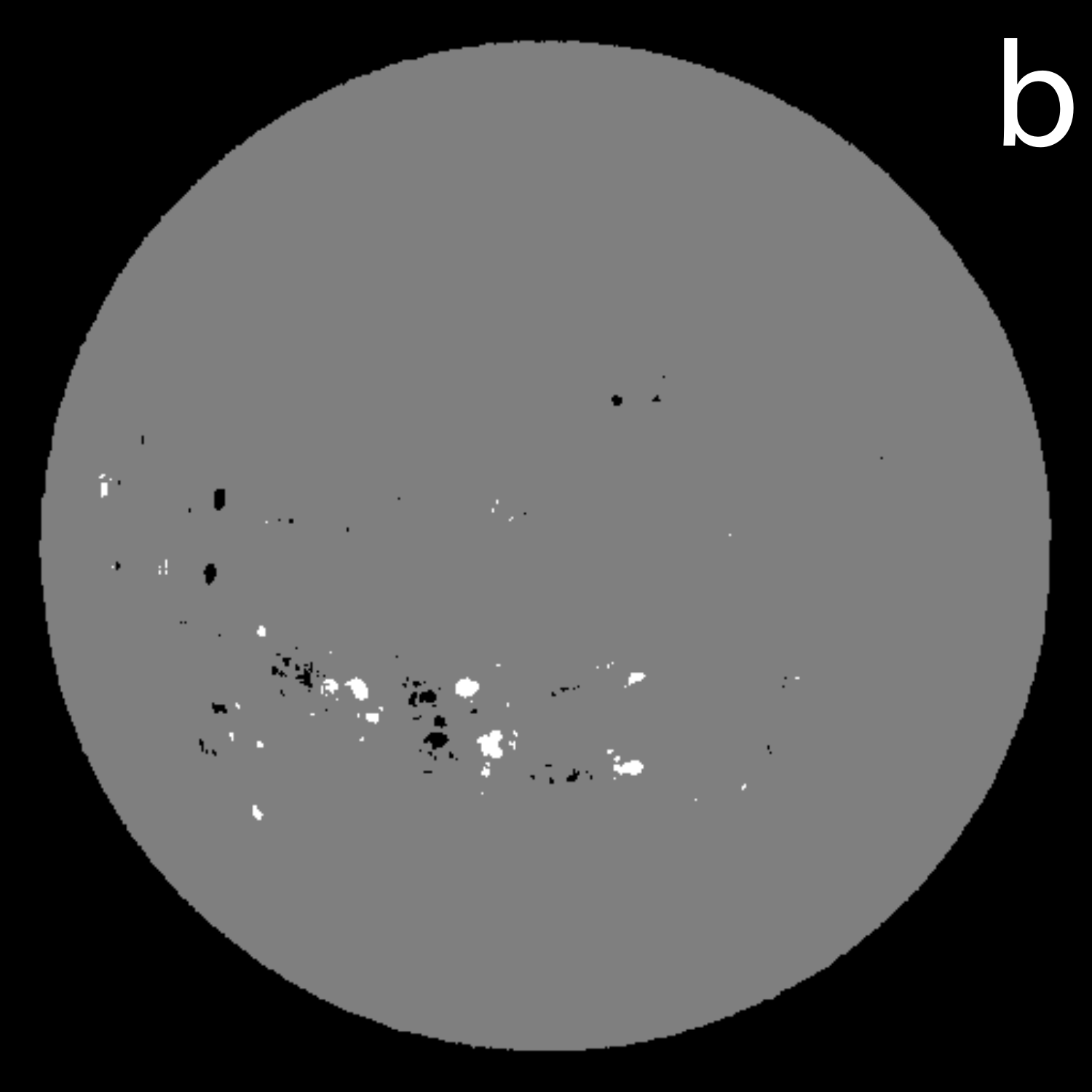}&
\includegraphics[width=0.28\textwidth]{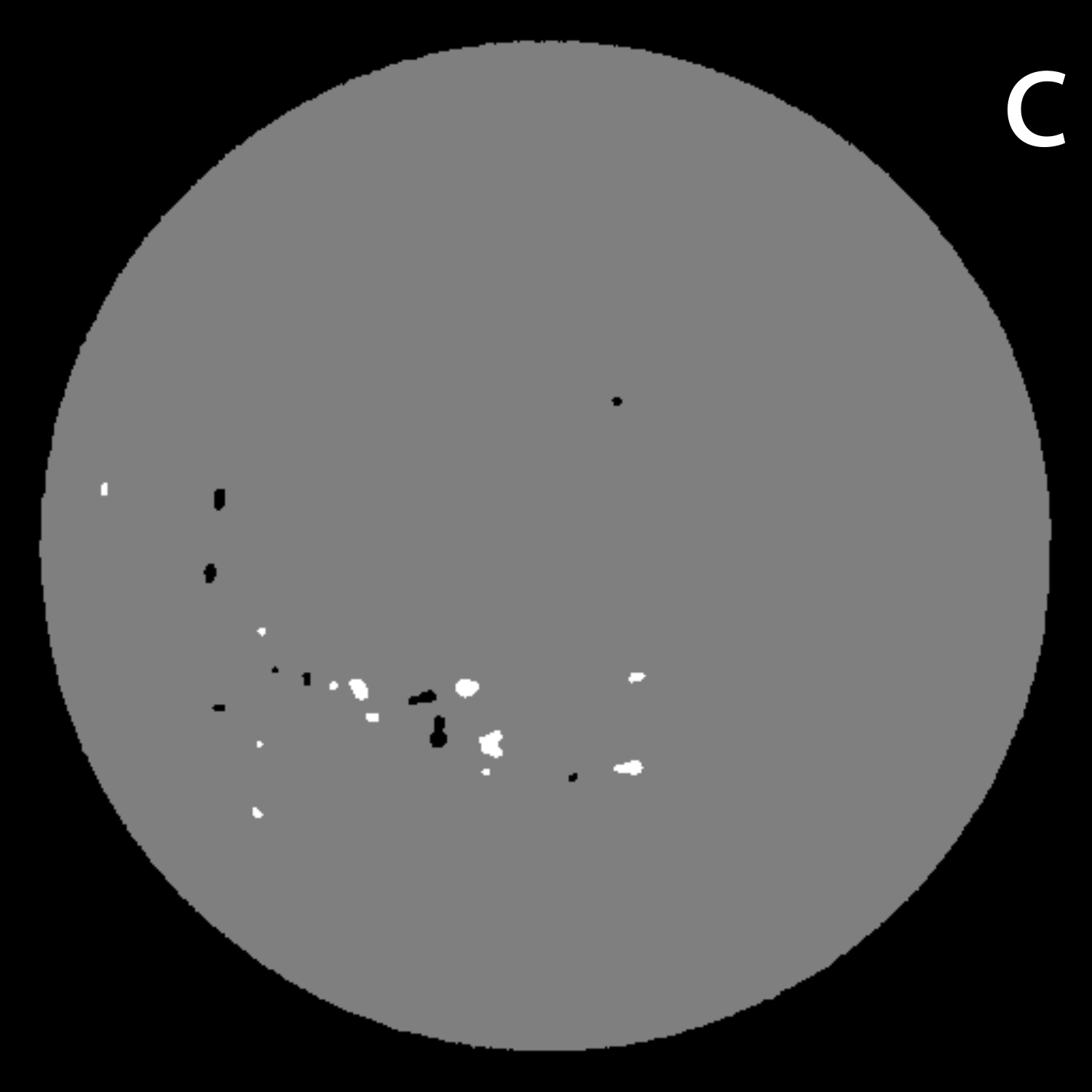}\\
\vspace*{-2.75cm}
\includegraphics[width=0.28\textwidth]{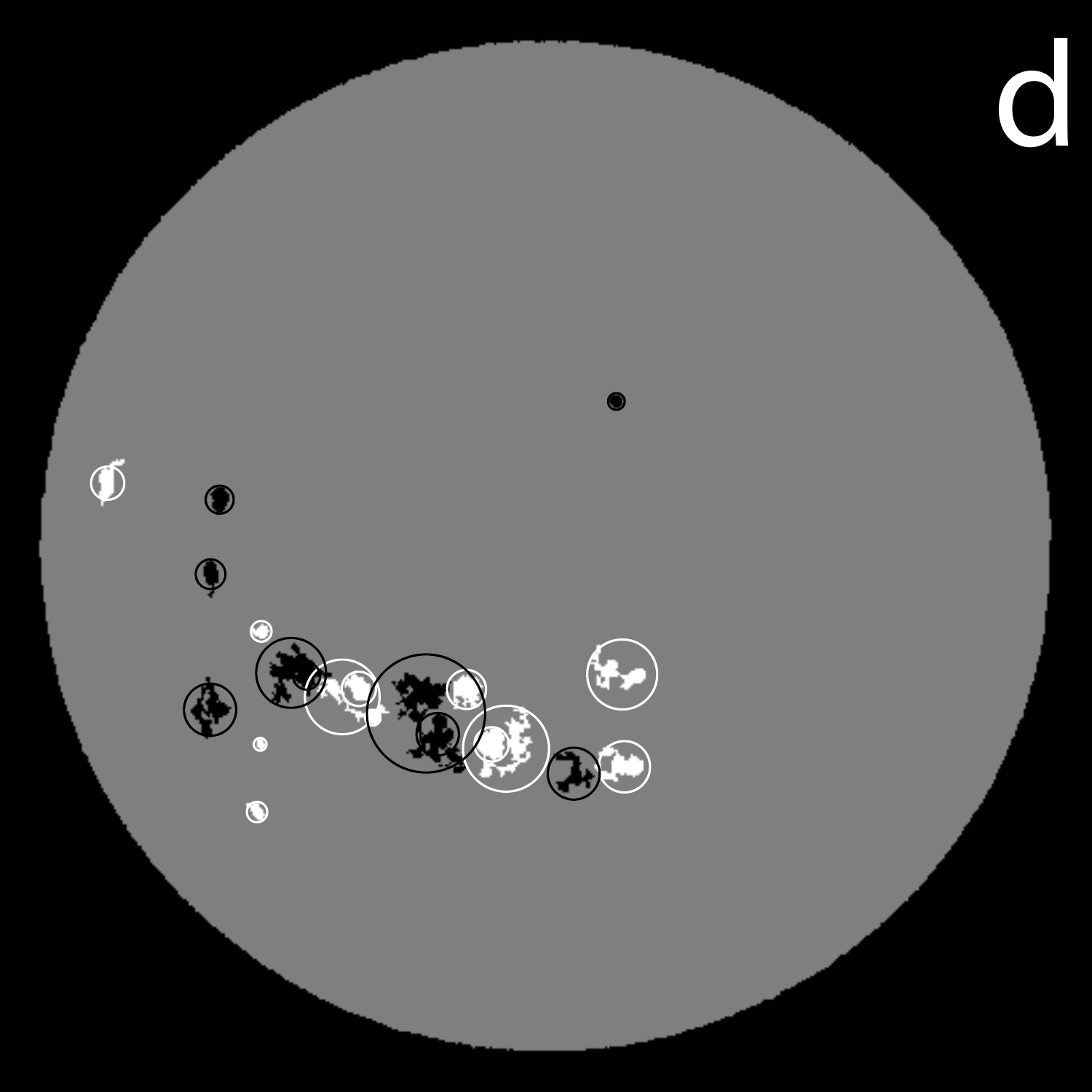}&
\includegraphics[width=0.28\textwidth]{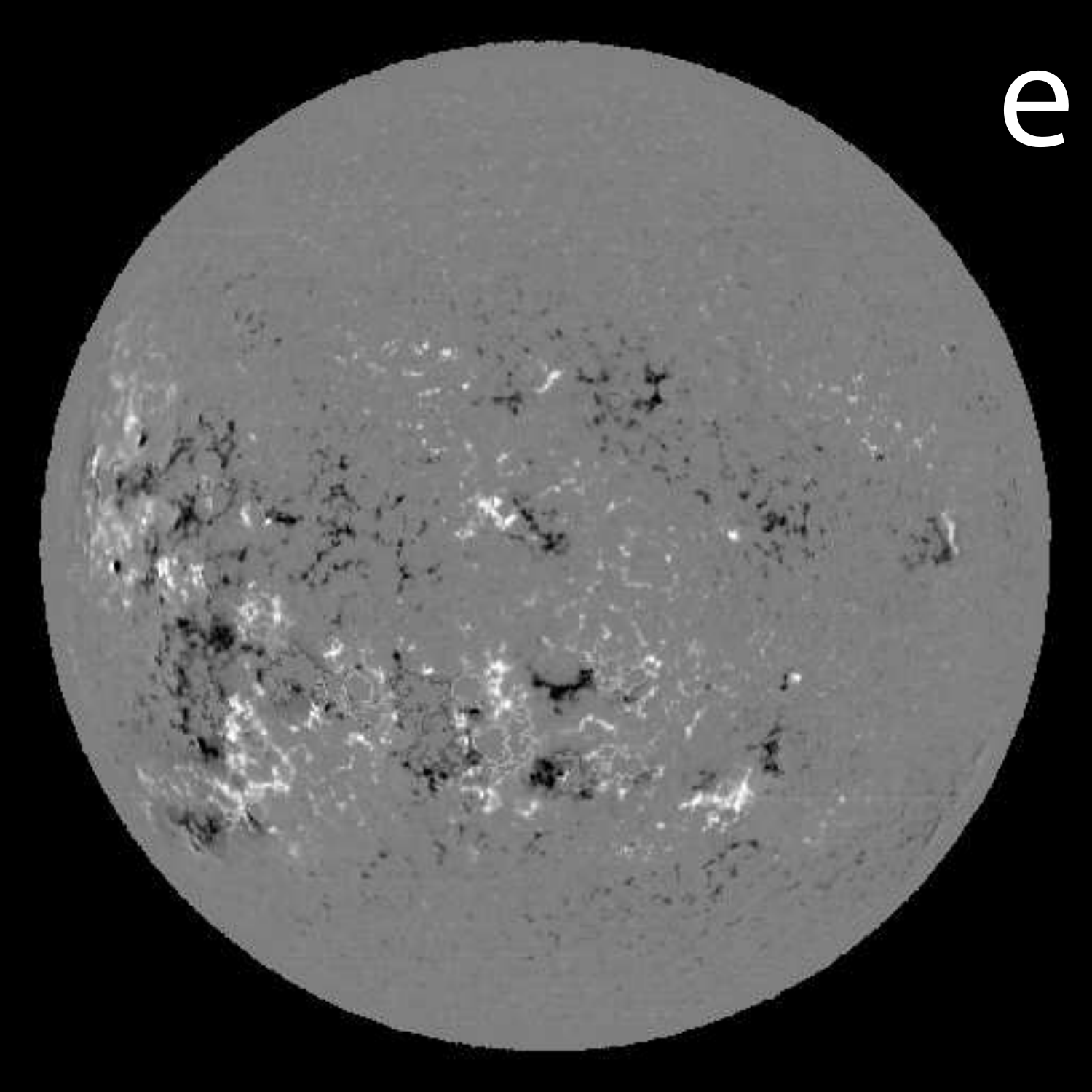}&
\includegraphics[width=0.28\textwidth]{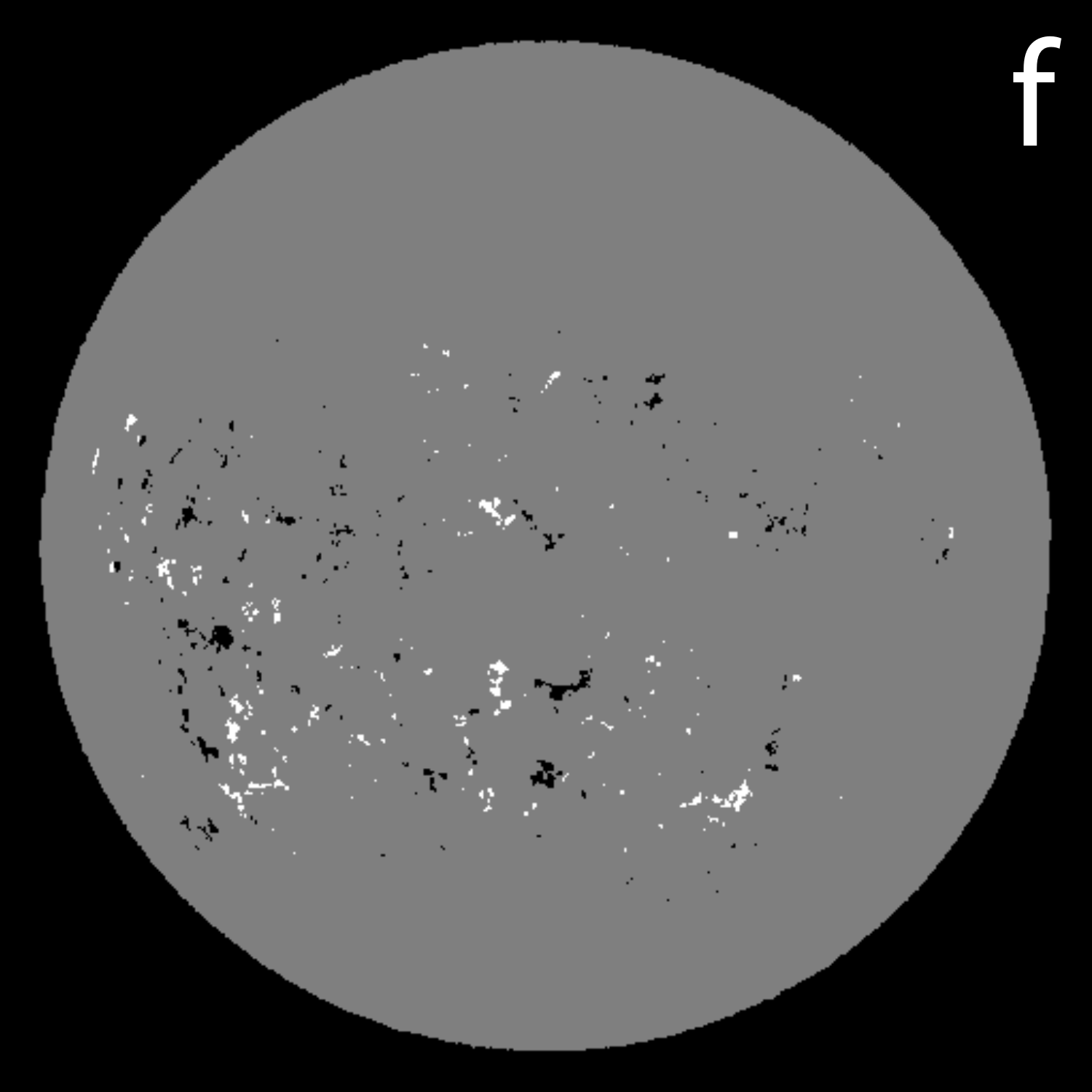}\\
\includegraphics[width=0.28\textwidth]{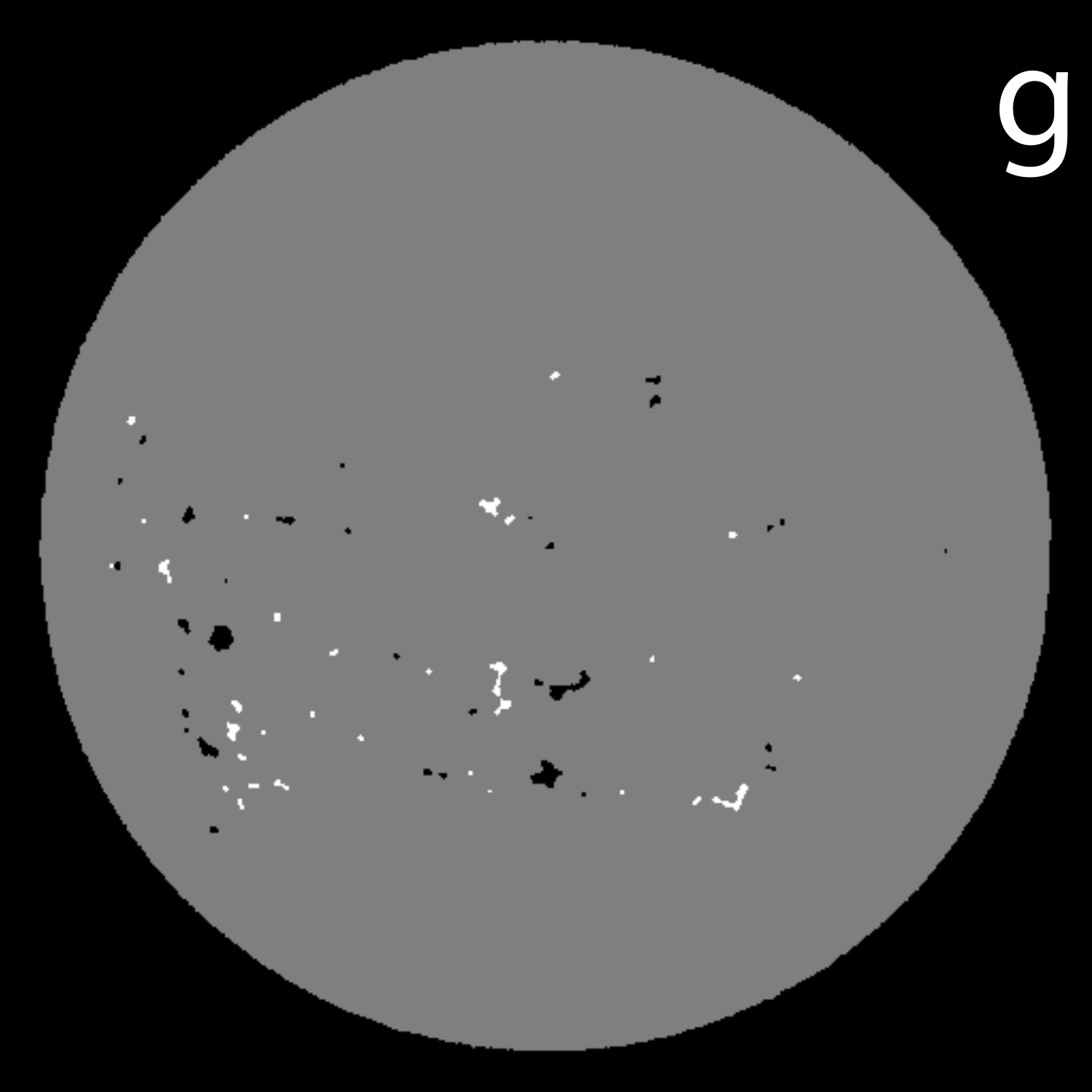}&
\includegraphics[width=0.28\textwidth]{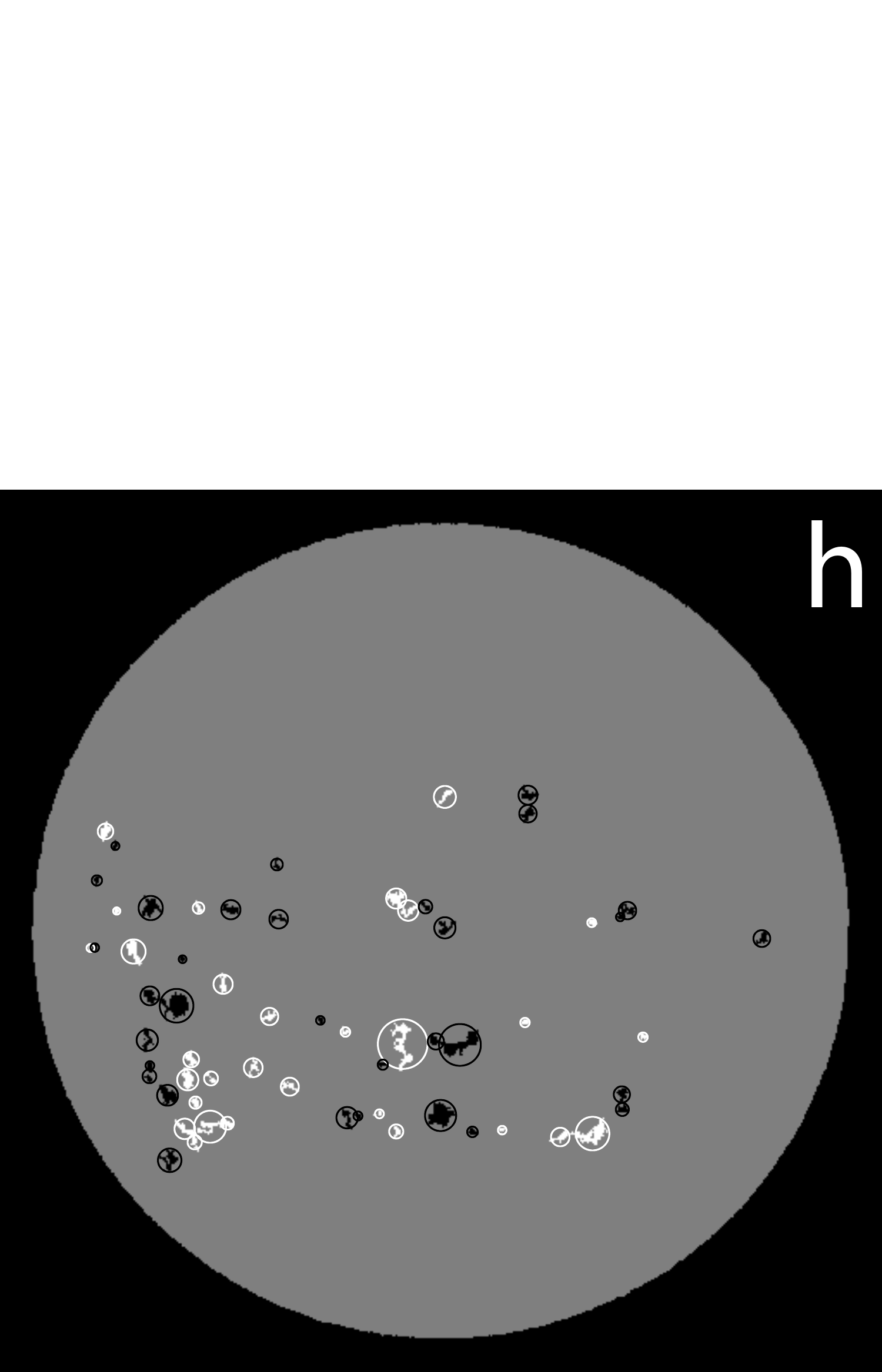}&
\includegraphics[width=0.28\textwidth]{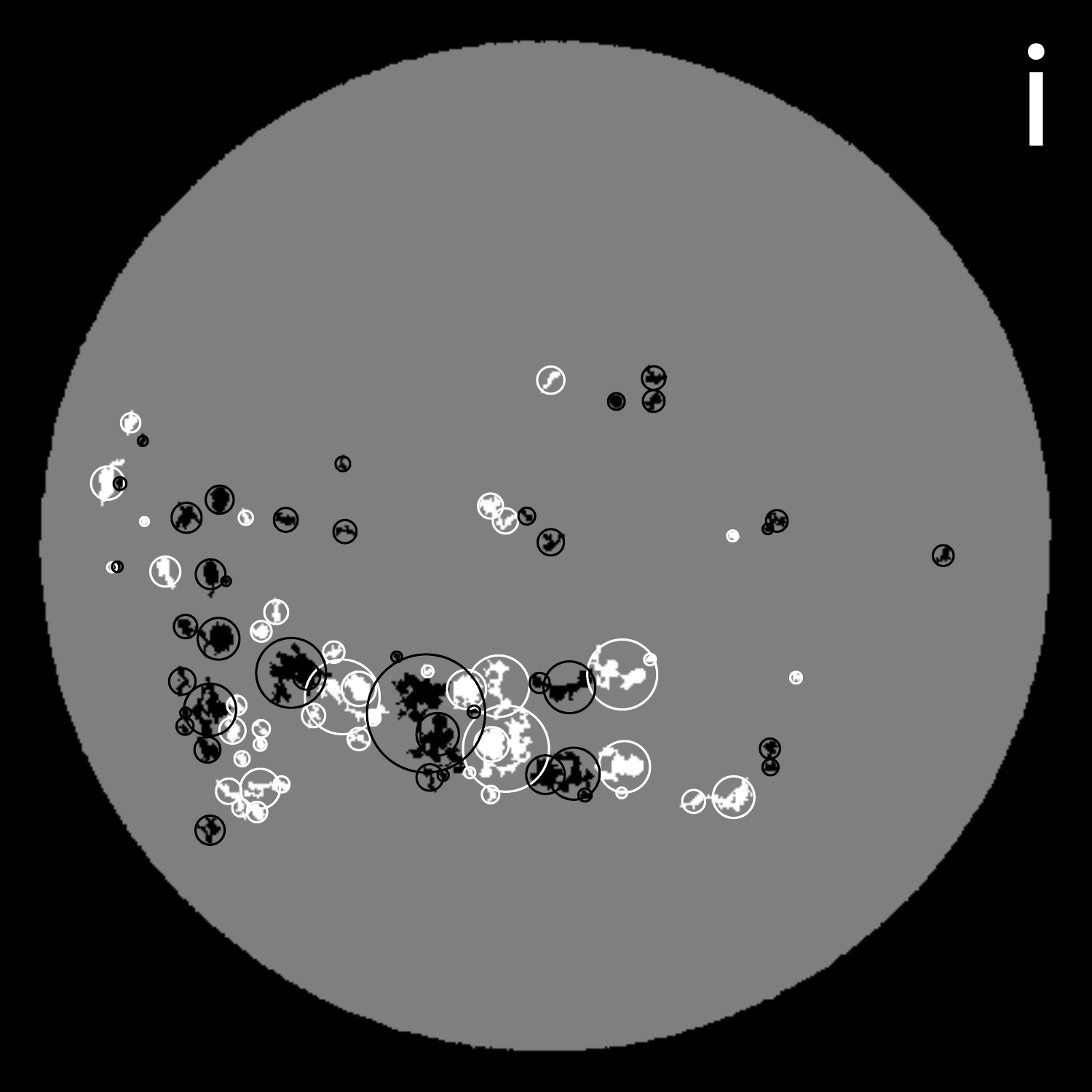}\\
\includegraphics[width=0.28\textwidth]{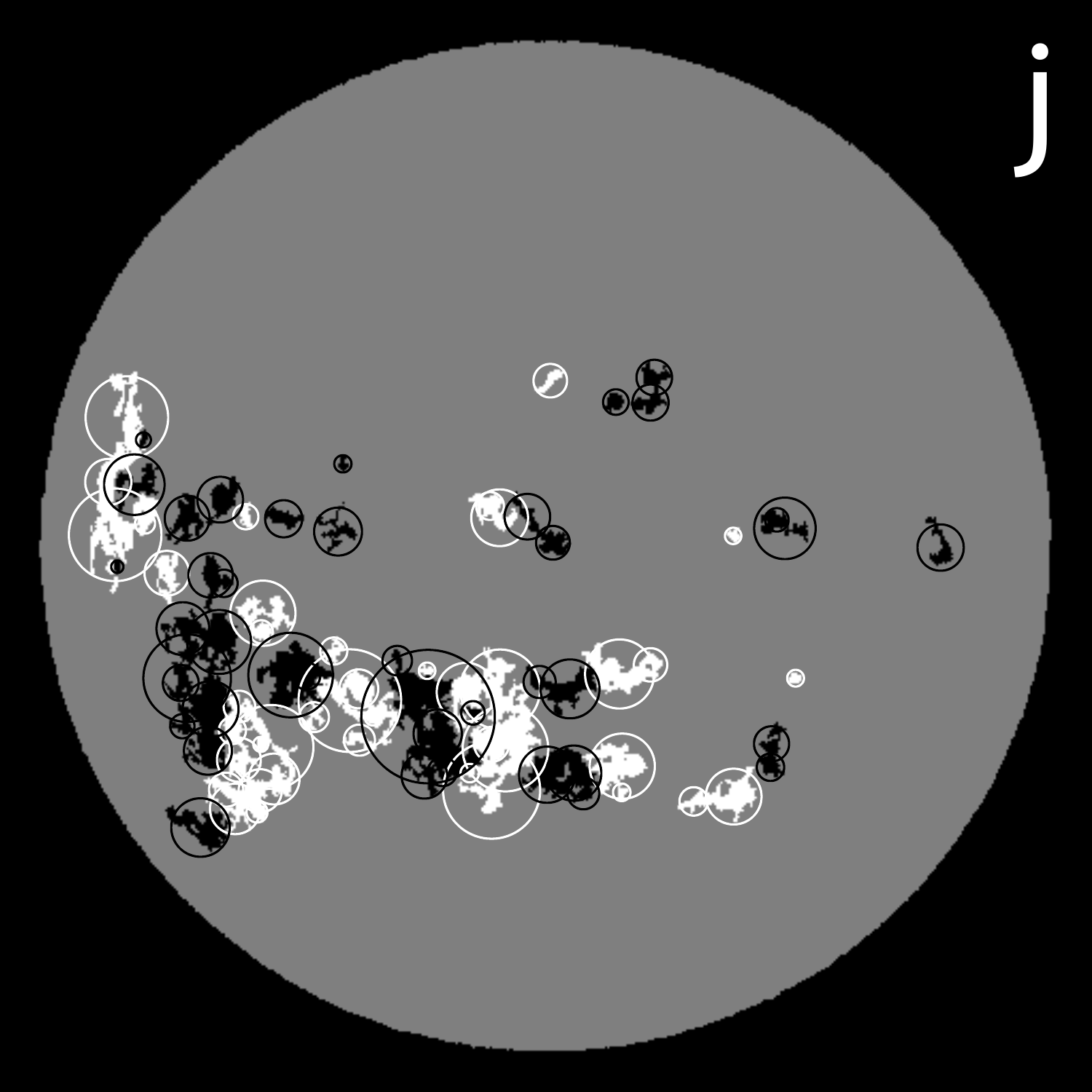}&
\includegraphics[width=0.28\textwidth]{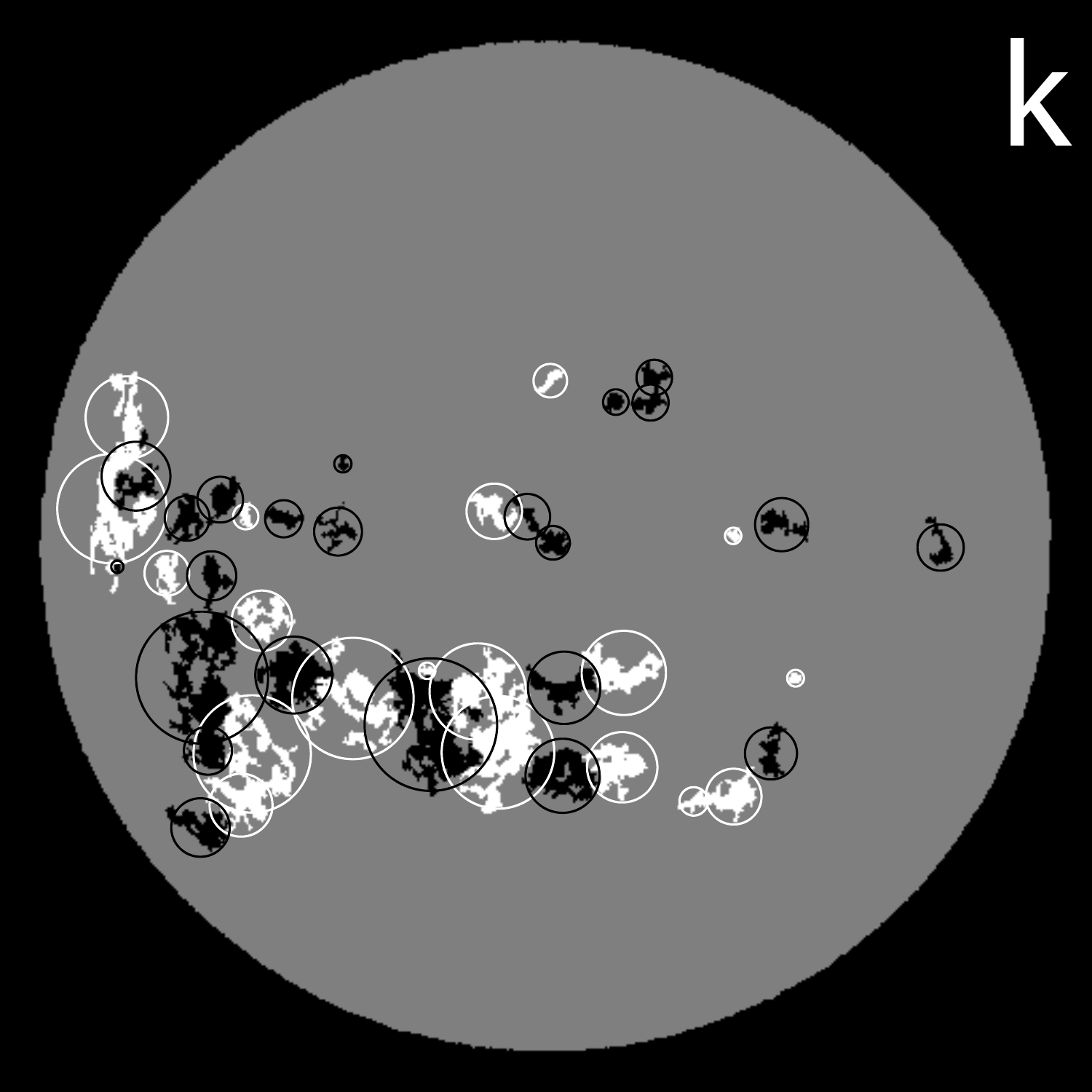}&
\includegraphics[width=0.28\textwidth]{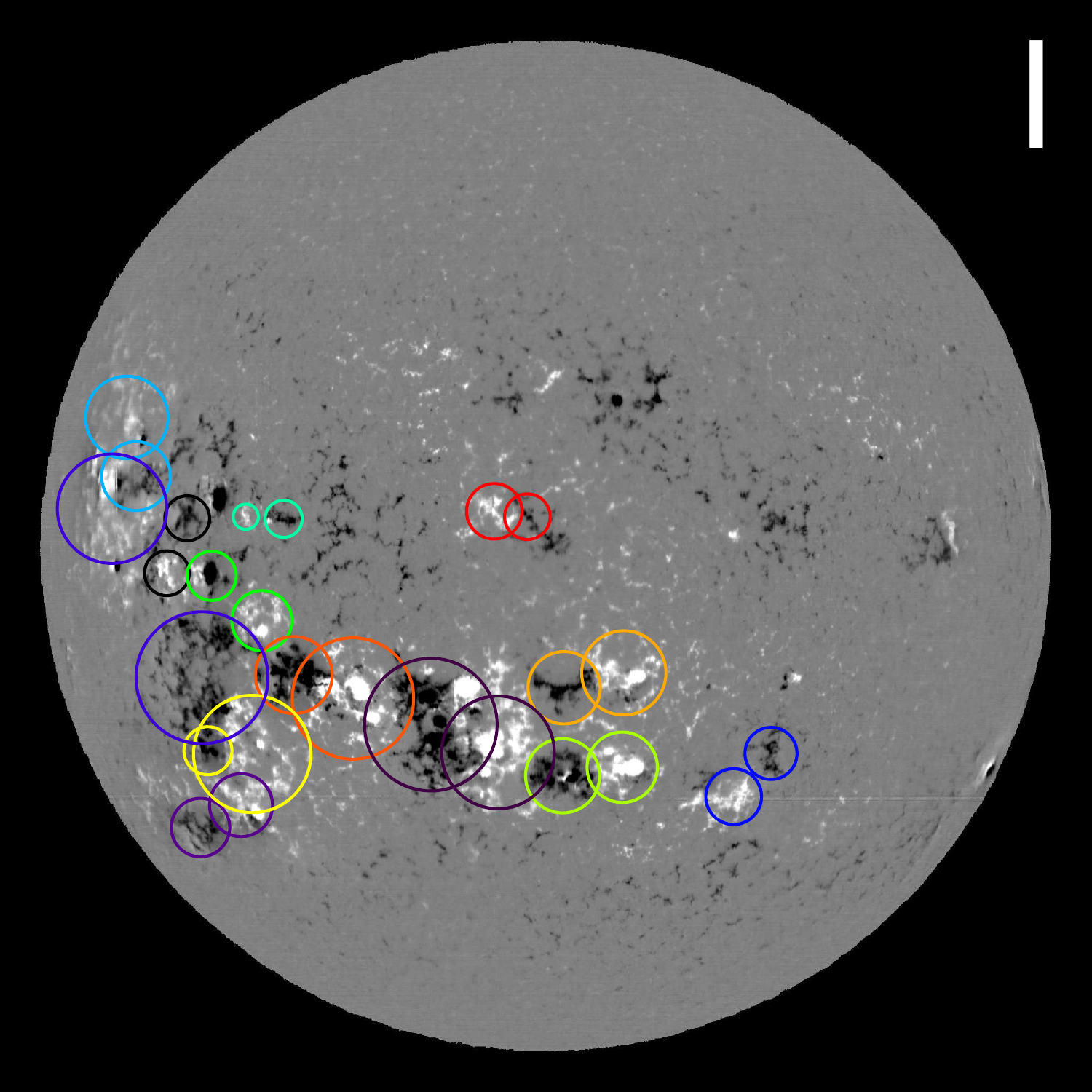}
\end{tabular}
\caption{BARD Algorithm. (a) Input KPVT/512 magnetogram taken on 1991-Aug-18. (b) Kernel pixels after applying a 400G threshold. (c) Kernel pixels after morphological opening operation. (d) Detected regions after growing kernels to encompass all adjacent pixels above a 150G threshold. (e) Magnetogram after setting the detected regions to zero. (f) Kernel pixels after applying a 200G threshold. (g) Kernel pixels after morphological opening operation. (h) Detected regions after growing kernels to encompass all adjacent pixels above a 150G threshold. (i) Detected regions found with both thresholds. (j) Detected regions after growing kernels to encompass all adjacent pixels above a 50G threshold. (k) Detected regions after merging regions that are too close together. (l) Detected bipolar magnetic regions after using pairing module.}
\label{fig_BARD}
\end{figure*}

\begin{figure*}[!t]
\centering
\includegraphics[width=0.43\textwidth]{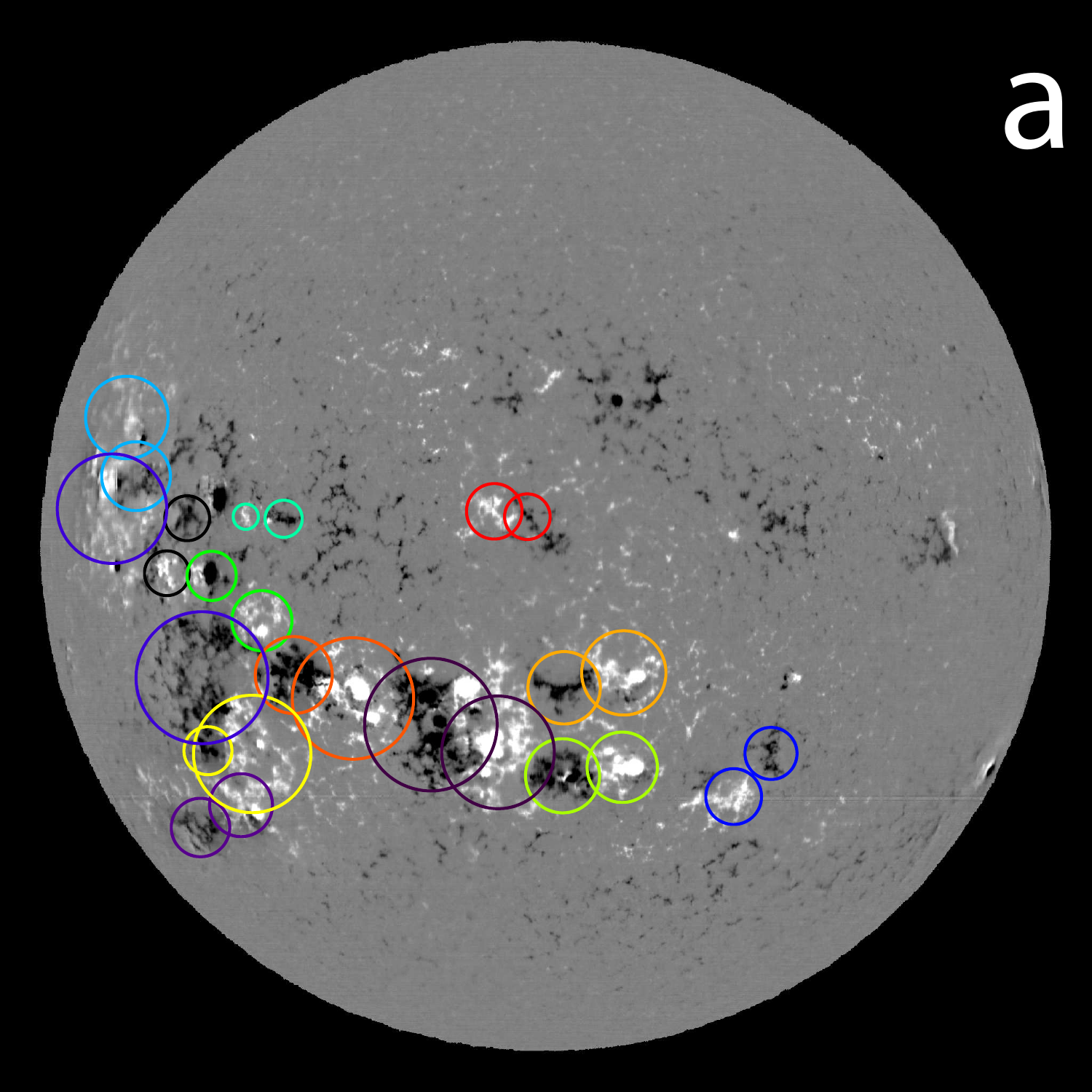}
\includegraphics[width=0.43\textwidth]{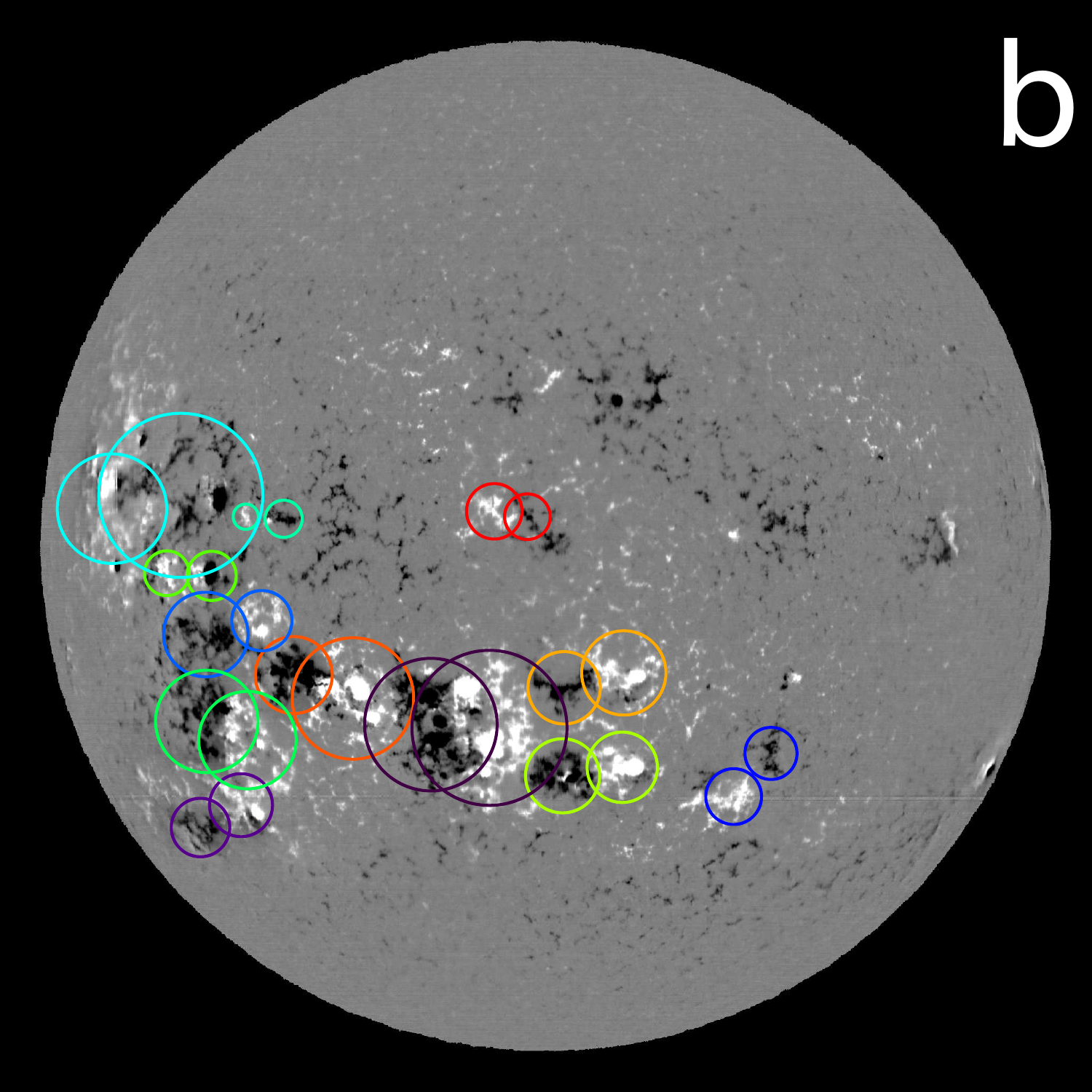}\\
\includegraphics[width=0.48\textwidth]{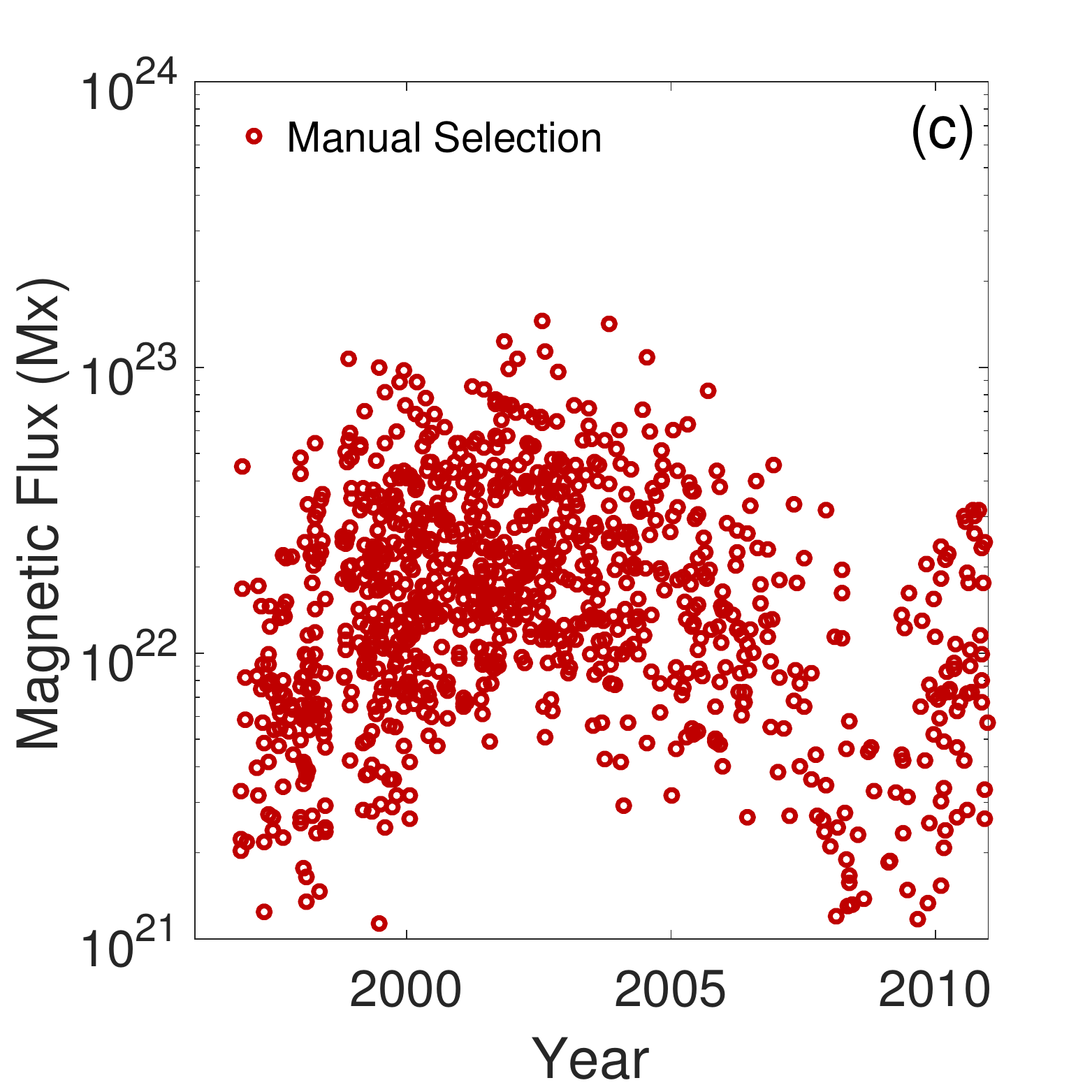}
\includegraphics[width=0.48\textwidth]{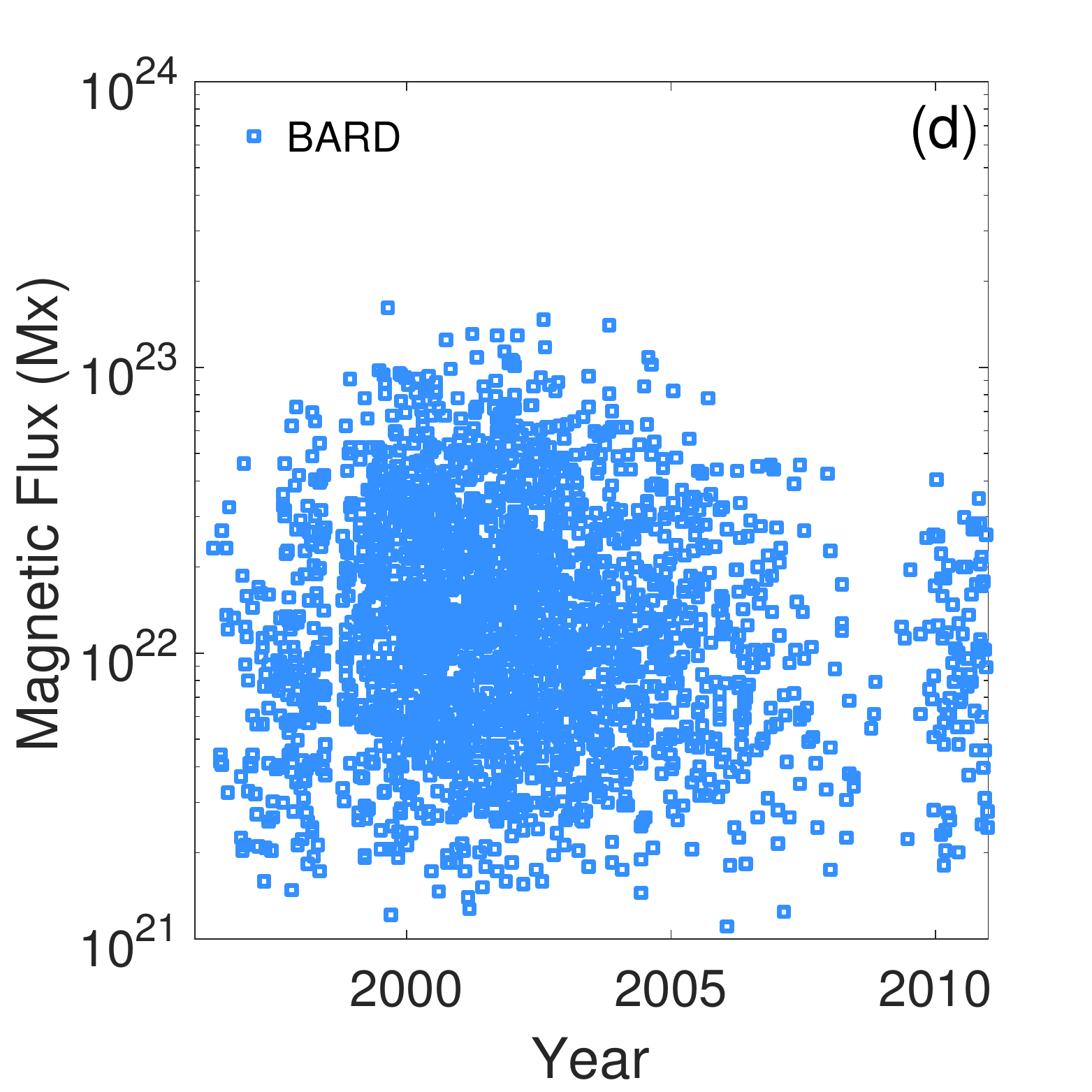}
\caption{Man vs.\ Machine. The top panels show a comparison between a fully automatic detection (a) of BMRs executed on KPVT/512 magnetogram taken on 1991-Aug-18 and a human vetted detection that involves the fragmentation of large regions in smaller ones and the re-pairing of negative and positive regions.  BMRs that retain the same coloring across panels were not touched by the human observer. The bottom panels show total unsigned flux for each uniquely detected BMR at maximum development for the SOHO/MDI mission using a logarithmic scale. (c) Results using manual detection. (d) Results using the BARD algorithm.}
\label{fig_Hm_vs_Mch}
\end{figure*}

\section{The Bipolar Active Region Detection (BARD) code}\label{sec_BARD}

The BARD code is based on the algorithm of Zhang, Wang and Liu\cite{zhang-wang-liu2010}, which we will refer to as ZWL from now on.  ZWL was originally designed for the detection of ARs and, as it is the case for most AR detection algorithms, it operated on unsigned magnetic field.  The main differences between BARD and ZWL are:
\begin{itemize}
  \item BARD treats positive and negative polarities separately
  \item BARD uses multiple thresholds of detection to ensure the proper capture of both large and small positive and negative regions (a single detection merges many small regions into neighboring large ones).
  \item BARD includes a module that merges regions of the same sign to give rise to larger objects.
  \item BARD has a pairing module that links positive and negative regions to form BMRs.
\end{itemize}

\subsection{BARD algorithmic framework}

Figure \ref{fig_BARD} illustrates the main steps of BARD's algorithmic framework starting with an input magnetogram (Fig.~\ref{fig_BARD}-a):
\begin{enumerate}
  \item Find all pixels above a threshold of detection $k_{th1}$ (Fig.~\ref{fig_BARD}-b).
  \item Apply a morphological opening operation using erosion size $e_{sz}$ and dilation size $d_{sz}$ to clean small cores out of the set (Fig.~\ref{fig_BARD}-c).
  \item Grow kernels to encompass all adjacent pixels above a threshold $r_{th1}$ (Fig.~\ref{fig_BARD}-d).
  \item Remove detected regions from magnetogram by setting magnetic field to zero (Fig.~\ref{fig_BARD}-e).
  \item Find all pixels above a second threshold of detection $k_{th2}$ (Fig.~\ref{fig_BARD}-f).
  \item Apply a morphological opening operation using erosion size $e_{sz}$ and dilation size $d_{sz}$ to clean small cores out of the set (Fig.~\ref{fig_BARD}-g).
  \item Growing kernels to encompass all adjacent pixels above a threshold $r_{th1}$ (Fig.~\ref{fig_BARD}-h).
  \item Combine the results of step 3 and step 7 (Fig.~\ref{fig_BARD}-i).
  \item Grow all regions to encompass all adjacent pixels above a lower threshold $r_{th2}$ (Fig.~\ref{fig_BARD}-j).
  \item Merge positive and negative regions based on a combination of size and proximity (Fig.~\ref{fig_BARD}-k).  The decision to merge two regions is based on the intersection between circles with radius equal to the average radius of each region. If there is at least $o_{lim}$ proportional area overlap, the regions are merged.
  \item Pair positive and negative regions into BMRs. First priority is given to matches between the current frame's regions and BMRs detected on the previous magnetogram. The remaining positive and negative regions are paired by minimizing a function that depends on distance between positive and negative regions, size difference, and flux balance (Fig.~\ref{fig_BARD}-l).
\end{enumerate}

\begin{table}[!t]
\renewcommand{\arraystretch}{1.5}
\caption{Instrument-Specific Positive and Negative Region Detection Parameters}
\label{Tab_Params}
\centering
\begin{tabular*}{0.48\textwidth}{@{\extracolsep{\fill}} l c c c c c c c}
          & $k_{th1}$ & $k_{th2}$ & $e_{sz}$ & $d_{sz}$ & $o_{lim}$\\
\hline
KPVT/512  & 400G      & 200G      & 10px     & 20px     & 0.2\\
KPVT/SPMG & 400G      & 200G      & 10px     & 20px     & 0.2\\
SOHO/MDI  & 500G      & 275G      & 9px      & 18px     & 0.2\\
SDO/HMI   & 200G      & 200G      & 10px     & 30px     & 0.0
\end{tabular*}
\end{table}

Most of the parameters that control the detection of positive and negative regions (steps 1 through 10) are instrument specific and can be found on Table \ref{Tab_Params}, the exceptions are $r_{th1}=$150G and $r_{th2}=$50G.  The function used to pair positive and negative regions is:
\begin{equation}\label{Eq_Pair}
  F(d,\Delta A, \Delta \Phi) = d^{a} \Delta A^{b} \Delta \Phi^{c},
\end{equation}
where $d$ is the distance between centroids in degrees, $\Delta A$ is the difference in area in cm$^2$, $\Delta \Phi$ is the flux imbalance in Mx, and $a=4$, $b=1$, and $c=1$.  These parameters are chosen empirically to obtain the best match between human and automatic detections performed on the same set of observations.

\begin{figure*}[!t]
\centering
\includegraphics[width=\textwidth]{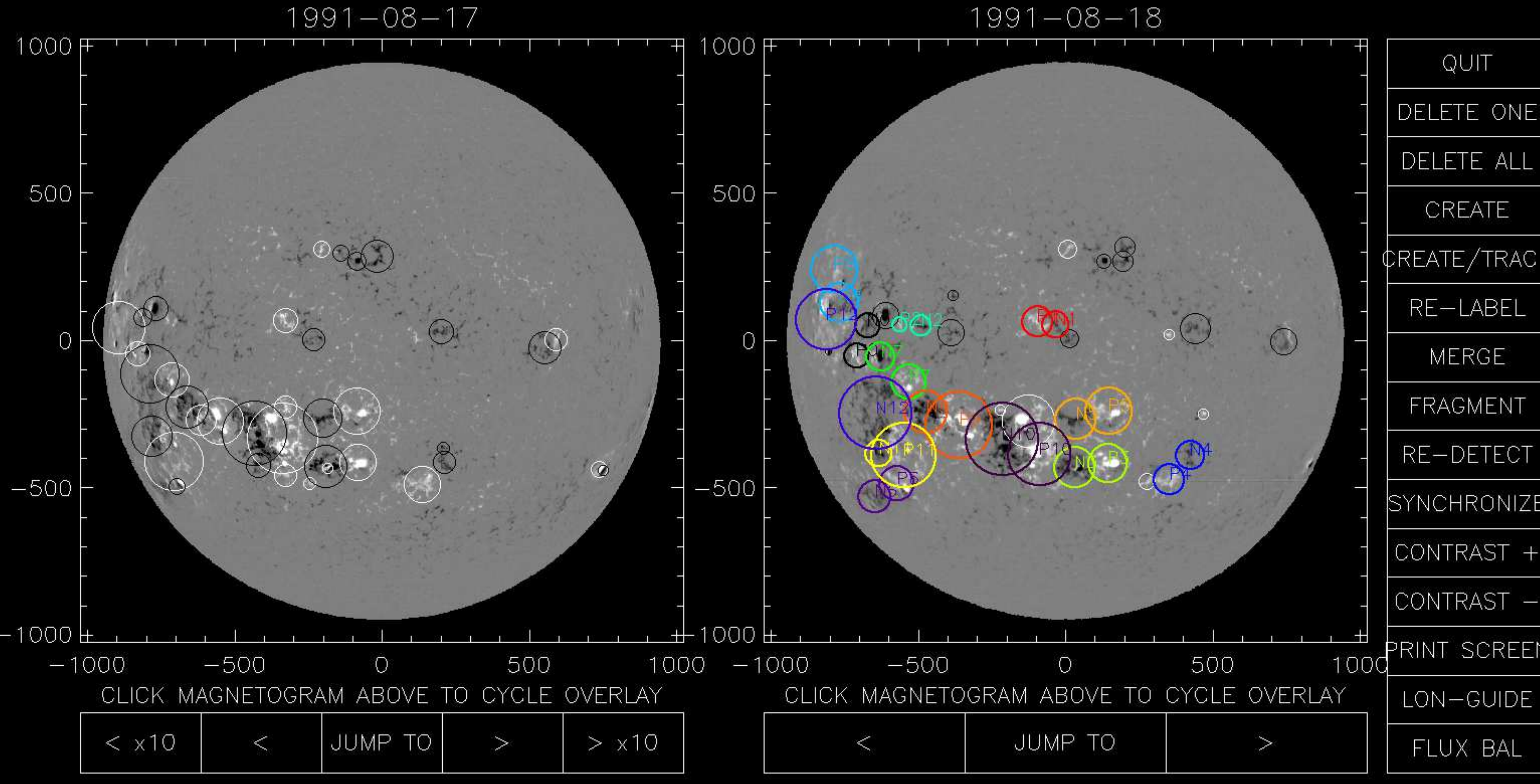}
\caption{Interface for BARD's human supervision module.  This interface was used to transform Fig.~\ref{fig_Hm_vs_Mch}-a into Fig.~\ref{fig_Hm_vs_Mch}-b.}
\label{fig_Supervision}
\end{figure*}


\section{Human vs.\ Machine: The Best of Both Worlds}\label{sec_mam_mach}

As BARD grew in complexity and sophistication, we realized that regardless of our algorithmic complexity, from time to time there would always BMR be pairings in which we disagreed with the result of the algorithm.  This is illustrated in Figures \ref{fig_Hm_vs_Mch}-a \& b which show a side by side comparison of the output of the BARD code when applied to one of the most complex magnetograms of the last 40 years (Fig.~\ref{fig_Hm_vs_Mch}-a) vs.\ a human realization of the BMR pairing (Fig.~\ref{fig_Hm_vs_Mch}-b). In this most difficult case, about half of the detected BMRs needed to be re-paired (untouched BMRs retain their colors).

To test the other side of the automatic vs.\ manual spectrum, we did an exercise where we placed the entire burden of detection on a human observer.  We found that there is a clear limit in complexity beyond which a human observer begins to miss BMRs.  This can be seen very clearly in Figures \ref{fig_Hm_vs_Mch}-c \& d which show a side by side comparison of the logarithm of BMR flux for all detected objects by a human observer (Fig.~\ref{fig_Hm_vs_Mch}-c) and by the BARD code (Fig.~\ref{fig_Hm_vs_Mch}-d) on SOHO/MDI magnetograms.  There is a very clear deficit of smaller objects in the human detection as solar activity ramps up.

\subsection{Human Supervision of the BARD Code}

In order to take advantage of the strengths of automatic and manual forms of detection, we implemented a module, shown in Figure \ref{fig_Supervision}, that allows a human observer to interact with the output of the BARD code. Using this module, the human observer can change BMR pairing and labeling (the most common mistakes of the BARD code), while placing all the detection burden on the automatic component.  Additionally, in order to keep the catalog as homogeneous as possible and the supervision exercise trackable, we limited the cadence of observations to one magnetogram per day -- even for modern high cadence instruments like SOHO/MDI and SDO/HMI.

The great majority of human interaction is performed by disassociation/association of positive-negative region pairs, followed by a relabeling step ensuring that a given object is properly tracked in time.  However, sometimes the algorithm merges adjacent, but separate BMRs.  When this is the case, the human observer can fragment a region into smaller objects by reversing step 10 (Fig.~\ref{fig_BARD}-k).

After having gone through the exercise of supervised detection with three different observers and for different magnetogram sources, our estimated proportion of human intervention is around 10\% of all objects in our database.

\begin{figure*}[!t]
\centering
\includegraphics[width=\textwidth]{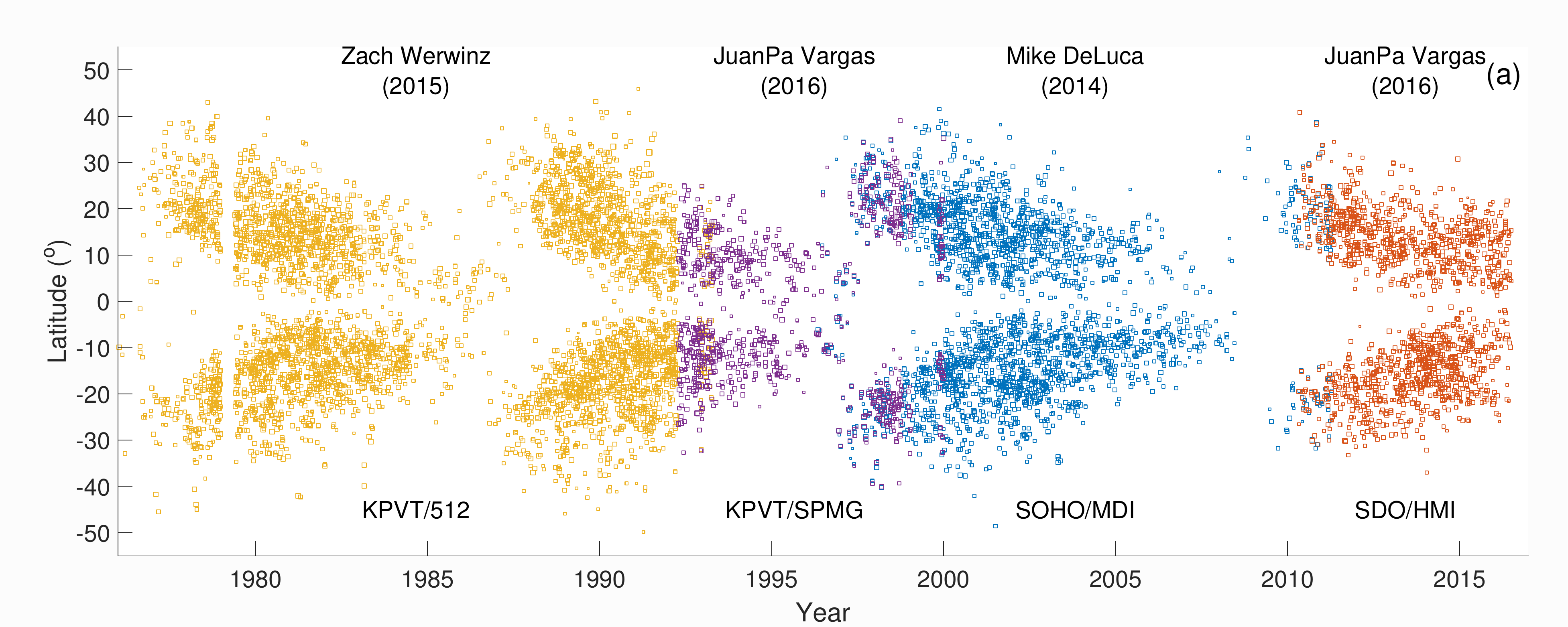}
\includegraphics[width=\textwidth]{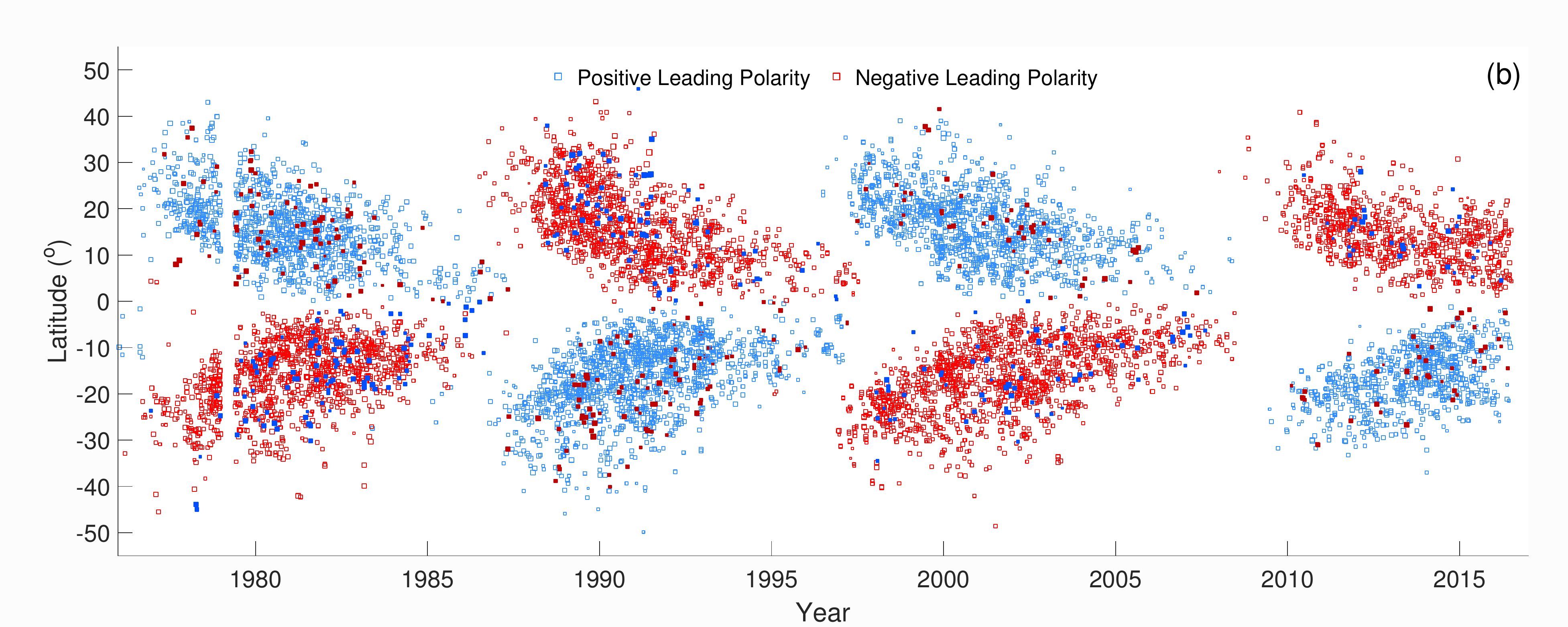}
\includegraphics[width=\textwidth]{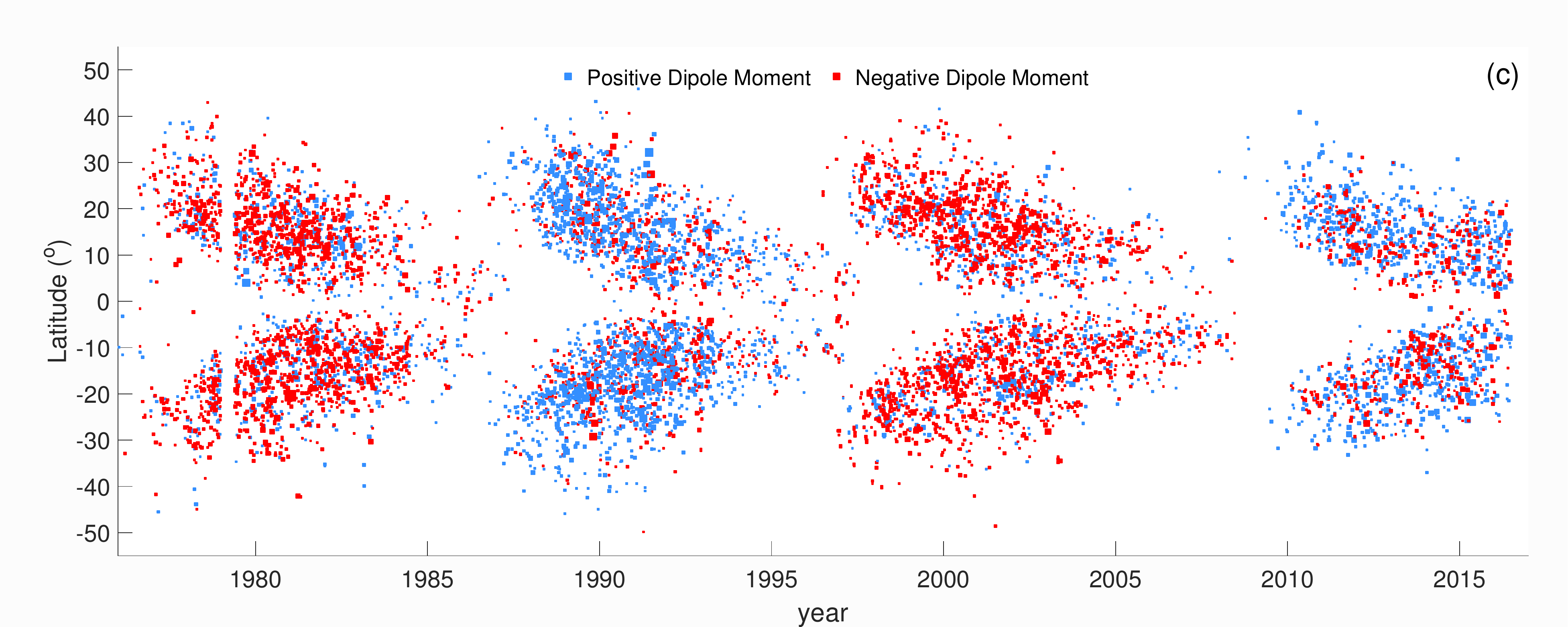}
\caption{ Butterfly diagram of each individually tracked object currently in the BARD catalog. (a) Marker color denotes the instrument that was used as data source. (b) Marker color denotes the sign of each BMR's leading polarity (solid markers indicate hemispheric exceptions to Hale's law). (c) Marker color denotes the sign of each BMR's contribution to the Sun's dipole moment.}
\label{fig_Bfly_students}
\end{figure*}

\section{Preliminary Results}\label{sec_results}

The assembly of the BARD catalog used magnetograms from four different instruments: the 512-channel of the Kitt Peak Vacuum Telescope (KPVT/512; 1976-1993), KPVT/SPMG (1992-1999), SOHO/MDI (1996-2011) and SDO/HMI (2010-2016).  Three different undergraduate research assistants worked over a period of three years (2014-2016) to complete it.

Figure \ref{fig_Bfly_students}-a shows a butterfly diagram in which every dot represents a unique BMR detected and tracked as it crosses the solar disk, colored according to the instrument it was measured on.  In spite of our data not yet being cross-calibrated, something on which we are working at the moment, it is reassuring to see agreement between instruments in terms of the time-latitude coordinates of the detected objects.

At the moment, the BARD catalog contains nearly 10,000 unique objects, all tracked across the disk, containing for each day of observation: a pixel mask of its positive and negative regions, positive and negative magnetic fluxes, latitude and longitude of each polarity's centroid, area covered by each polarity, and tilt.  This information is displayed as an example in Figure \ref{fig_Bfly_students}-b, which shows the systematic East-West hemispheric orientation of BMRs (also known as Hale's law\cite{hale-etal1919}). The combination of this systematic East-West orientation and the tendency of the leading (trailing) polarity of BMRs to be closer to the equator (pole), also known as Joy's law, results in the synergic buildup of a large-scale north-south magnetic field shown in Figure \ref{fig_Bfly_students}-c.

\section{Future Work: Building a Multi-Scale magnetic Catalog using BARD and SWAMIS}\label{sec_SWAMIS}

As mentioned in Section \ref{sec_results}, we purposely limited the scope of our catalog to maximize consistency and accuracy when detecting and characterizing BMRs.  The result is a catalog with lower cadence (one measurement per day) than SOHO/MDI (16 measurements per day) and SDO/HMI (one measurement every 10 minutes), as well as limited to the top three observable orders of magnitude in flux (i.e.\ $10^{20}-10^{22}$Mx) -- a limit imposed both by our chosen set of thresholds and the difficulty of determining bipolarity for smaller objects.  While this is sufficient to seed solar dynamo and surface flux transport models, as well as to characterize the large scale statistical properties of BMRs and their dependence of solar activity, there is immense value in building catalogs that take advantage of full instrumental cadence and scale.  The reason is that such a catalog will enable a wide array of statistical analyses that can shed light in the scaling and origin of surface magnetism\cite{harvey-zwaan1993,baumann-solanki2005,parnell-etal2009,munoz-etal2015a}, the nature of observable dynamo action\cite{lamb-etal2014,munoz-etal2015a}, dependence of surface magnetism on solar activity\cite{hagenaar-etal2008,munoz-etal2015b}, and characterization of the solar extended cycle\cite{harvey1992,mcintosh-etal2014}, among many other examples.

\subsection{SWAMIS: A multi-scale Feature Tracking Algorithm}

Starting development in 2001, SWAMIS\cite{deForest-etal2007} has evolved to be a truly multi-scale detection algorithm that has been used to study flux balance in solar surface magnetism\cite{lamb-etal2008,lamb-etal2010}, flux distribution of magnetic elements\cite{parnell-etal2009}, dissipation of surface magnetic elements\cite{lamb-etal2013}, and locality of surface dynamo action\cite{lamb-etal2014}.

\begin{figure}[!t]
\centering
\begin{tabular}{c}
\includegraphics[width=0.3\textwidth]{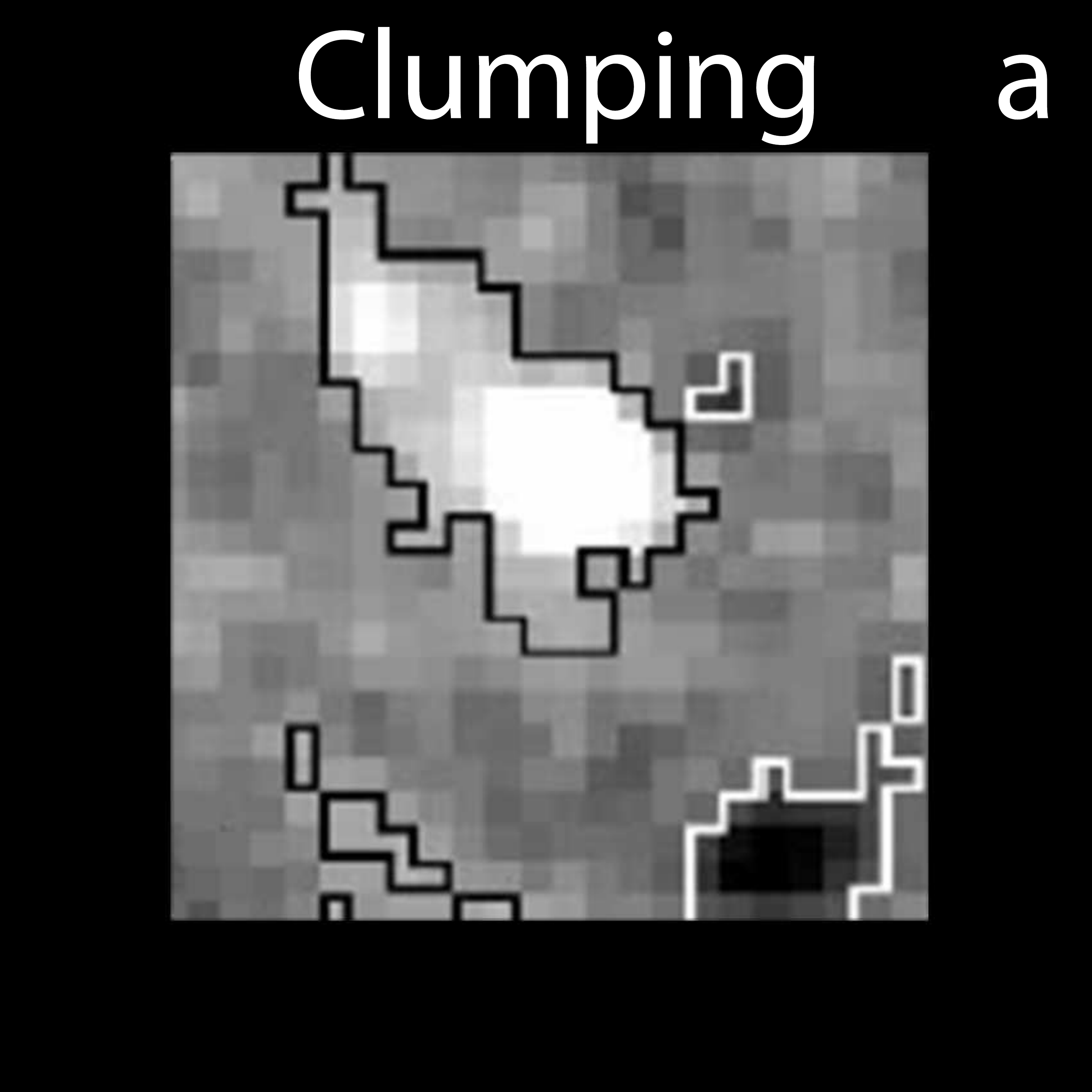}\\
\includegraphics[width=0.3\textwidth]{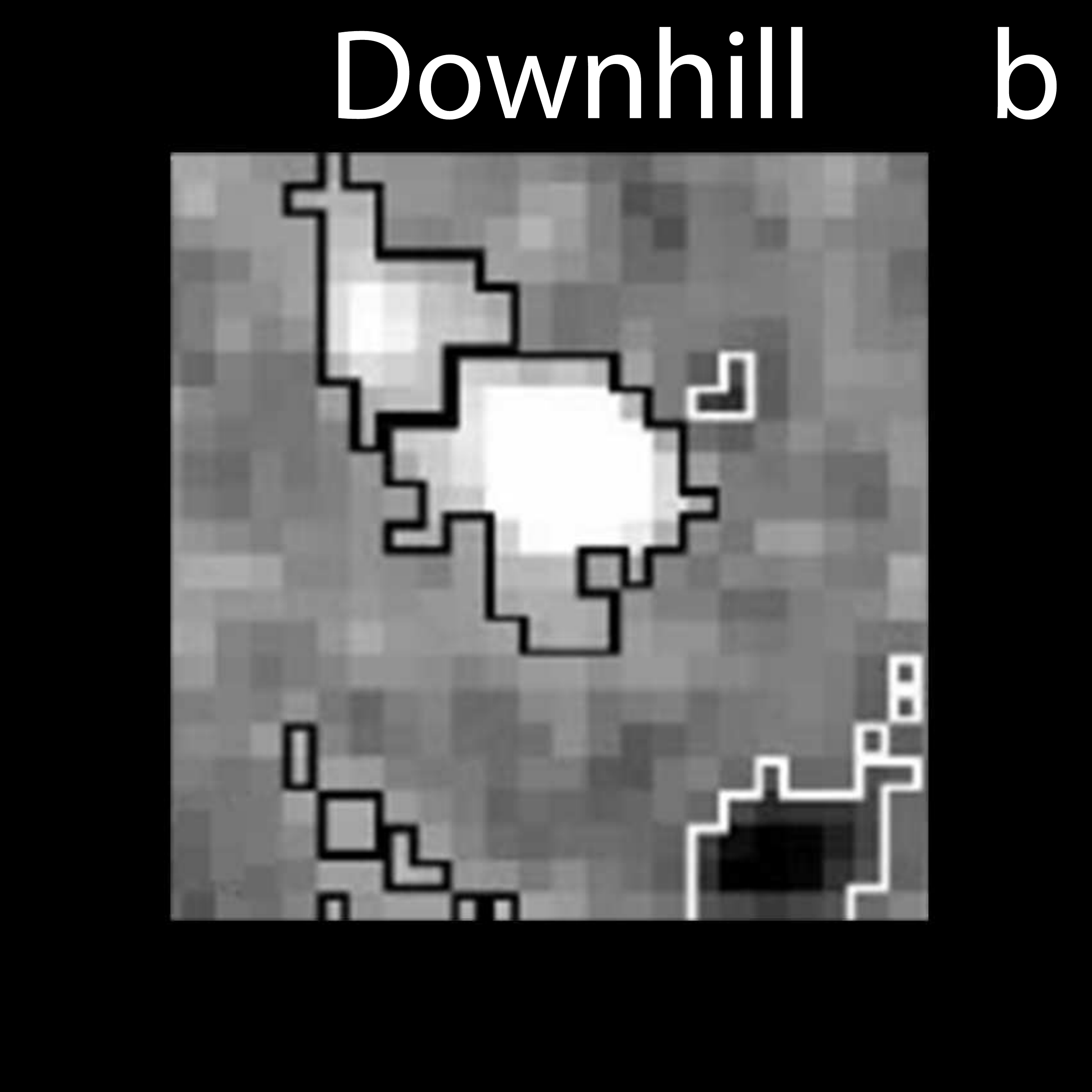}
\end{tabular}
\caption{Feature growing algorithms. Clumping (a) groups all adjacent pixels above a certain threshold of detection.  Downhill (b) seeds each region around a local maximum in flux and grows each element until their boundaries meet at saddle points. The current version of SWAMIS uses a hybrid method: clumping in areas of high average field strength, and downhill in areas of low average field strength. \textit{Adapted from\cite{deForest-etal2007}.}}
\label{fig_Algorithms}
\end{figure}

There are two main differences between BARD and SWAMIS. The first one is the algorithm used for fully determining the size of each feature.  As mentioned in Section \ref{sec_BARD}, BARD uses an algorithm that groups all adjacent pixels above a certain threshold of detection (also referred to as clumping; see Figure \ref{fig_Algorithms}-a).  This algorithm has a tendency to join large regions, which is not a problem for large BMRs, but becomes a significant problem for small flux scales.  To address this problem, SWAMIS combines the use of a clumping algorithm for large objects with a downhill algorithm for small objects(see Figure \ref{fig_Algorithms}-b), which seeds each region around a local maximum in flux and grows each feature until their boundaries meet at saddle points.

\begin{figure*}[!t]
\centering
\begin{tabular}{ccc}
\includegraphics[width=0.3\textwidth]{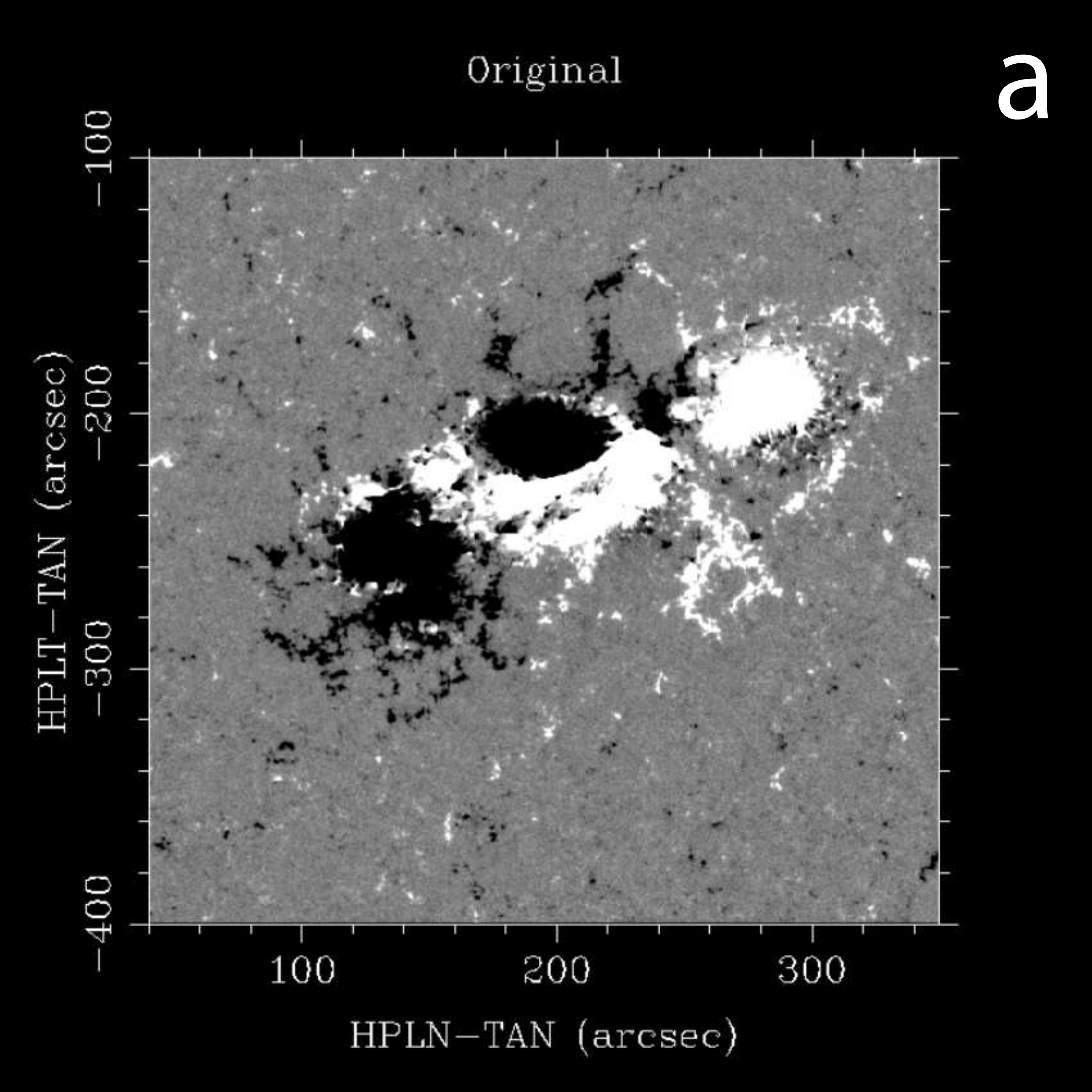} & \includegraphics[width=0.3\textwidth]{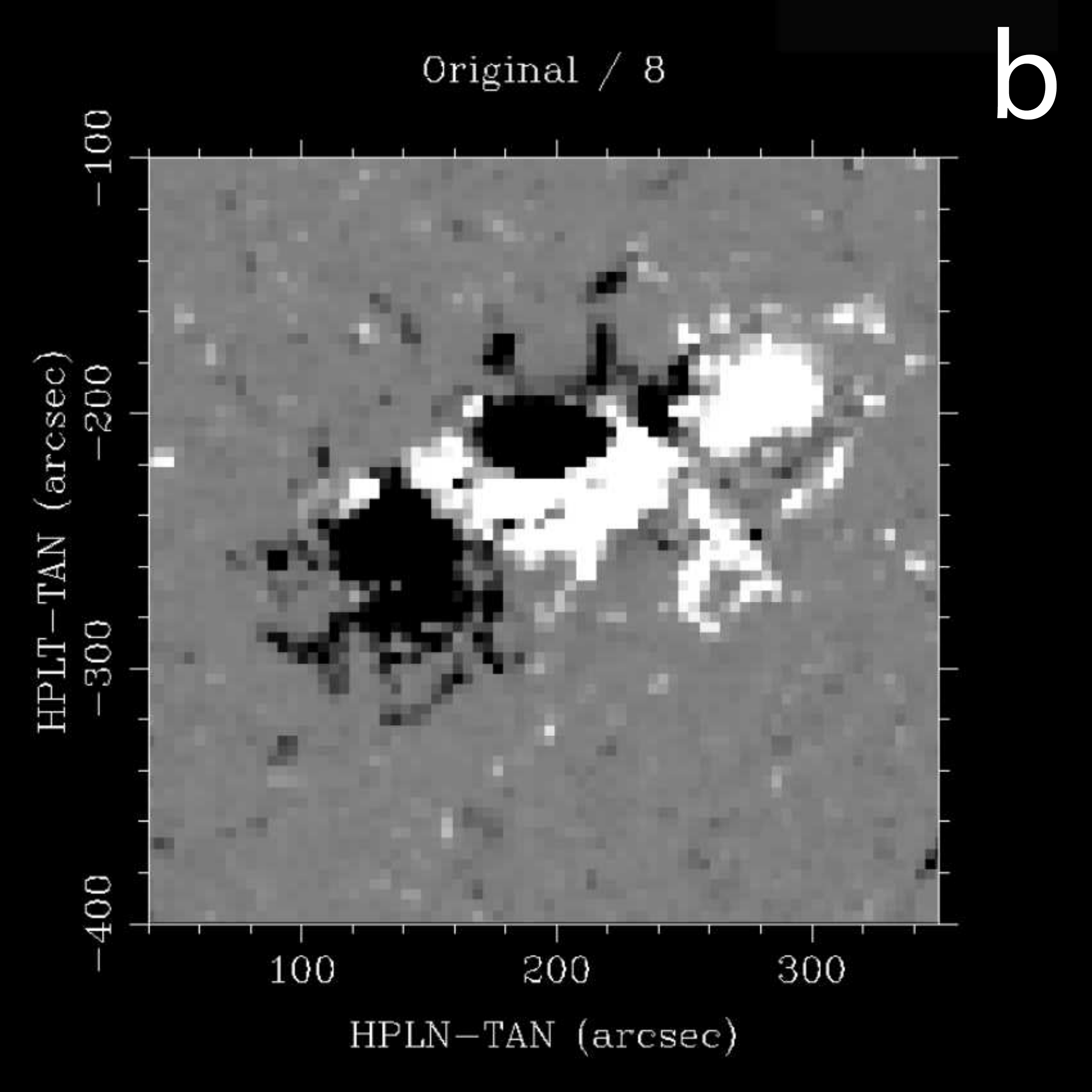} & \includegraphics[width=0.3\textwidth]{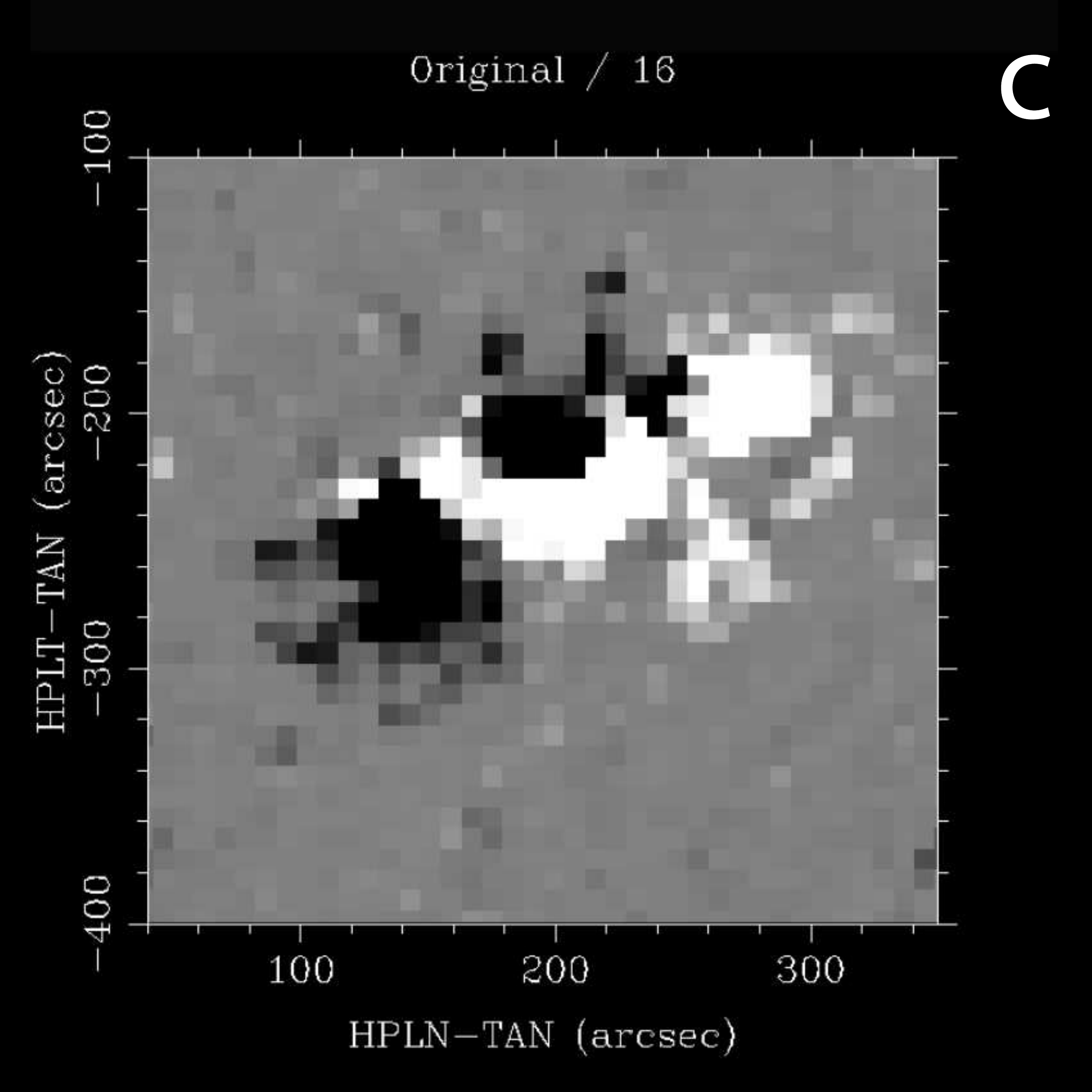}\\
                                                        & \includegraphics[width=0.3\textwidth]{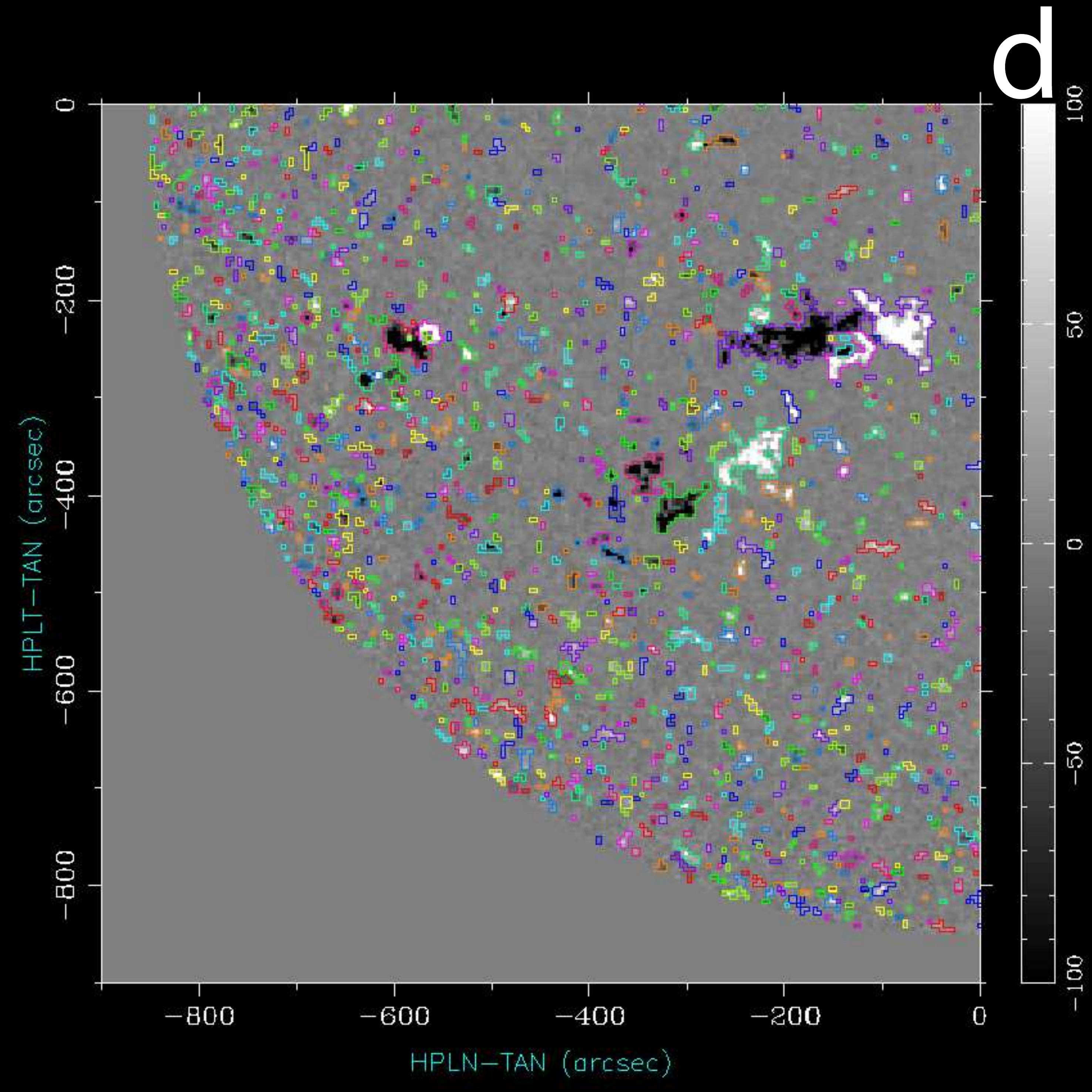} & \includegraphics[width=0.3\textwidth]{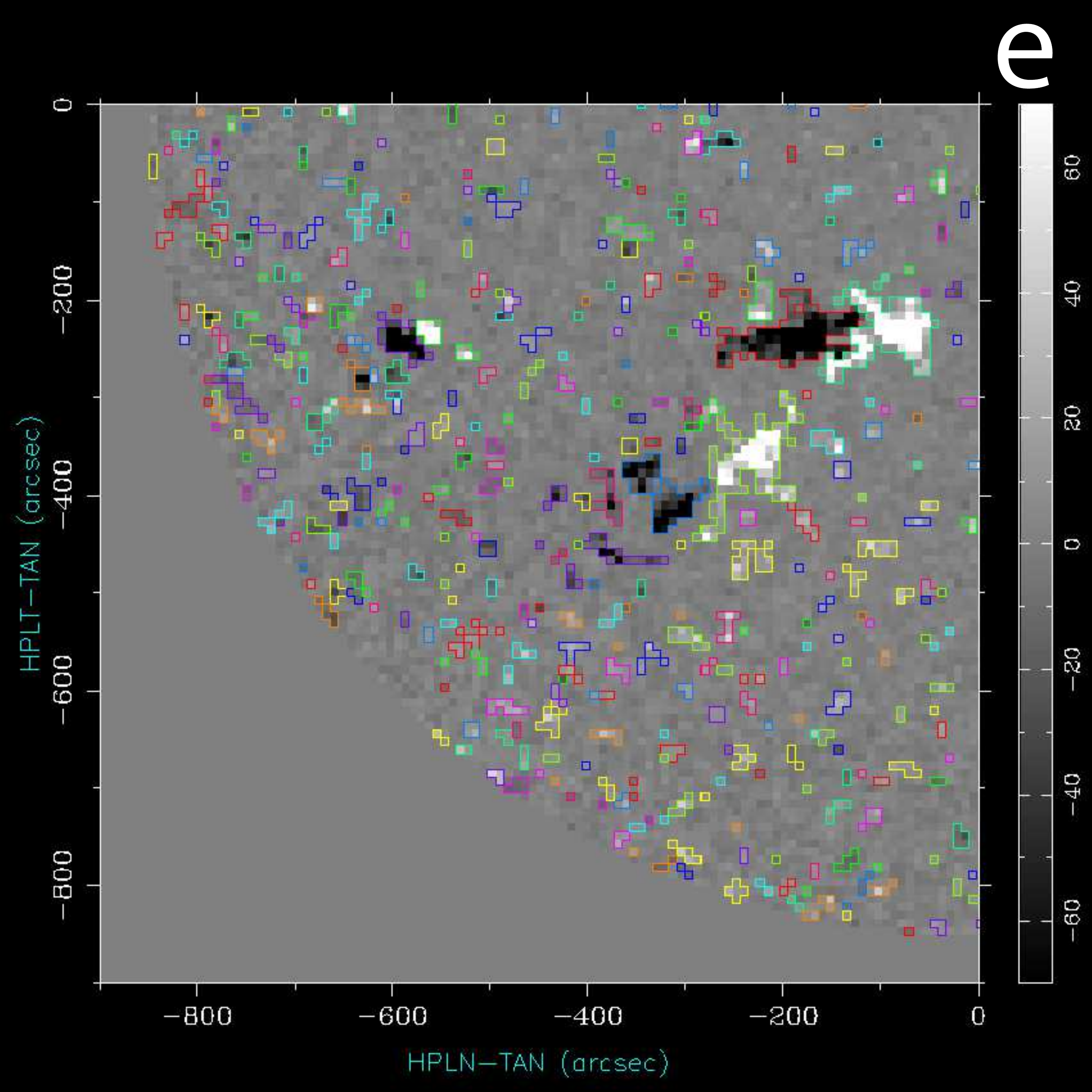}
\end{tabular}
\caption{(a) Original-resolution SDO/HMI magnetogram, degraded by a factor of 8 (b) and 16 (c). Feature detection and tracking of these reduced-resolution magnetograms allows for the association of smaller-scale features (d) with larger-scale features (e).}
\label{fig_SWAMIS}
\end{figure*}

The second main difference, which makes SWAMIS unique among magnetic feature tracking codes, is the application of the feature detection and tracking to resampled versions of the same original image.  This process, illustrated in Figure \ref{fig_SWAMIS}, determines the algorithmic make up of SWAMIS.  Each of the following steps are applied by SWAMIS to every magnetogram:

\begin{enumerate}
  \item Downscale the original image (Fig.~\ref{fig_SWAMIS}-a) to one half, one quarter, one eighth (Fig.~\ref{fig_SWAMIS}-b), and one sixteenth (Fig.~\ref{fig_SWAMIS}-c) of the original resolution.
  \item Identify which pixels in the images are in features by flagging pixels that are above a scale-dependent high threshold, and flagging adjacent (in space and time) pixels that are above a scale-dependent low threshold, and continuing the search until all such pixels have been flagged
  \item Apply the hybrid clumping/downhill feature identification method to segment each image into magnetic features (Figs.~\ref{fig_SWAMIS}-d\& e show the results for 1/8th and 1/16th of the original resolution), creating masks with feature IDs.
  \item If necessary based on image resolution and cadence, differentially rotate the masks of the previous magnetogram using the difference in time between current and previous frames.
  \item Use a dual-maximum flux-weighted overlap method\cite{deForest-etal2007} to identify feature to feature matches between the previous and current frames, updating current feature IDs if a match is found.
  \item Associate overlapping features across subsequent downsized resolutions.  This step is what makes SWAMIS a multi-scale code as it allows the user to connect each large-scale structure with its constituent smaller-scale features.
\end{enumerate}

\subsection{BARD \& SWAMIS: A Match Made in Heaven}

One of the important lessons that we have learned in making the BARD catalog is the difficulty of automatically pairing positive and negative magnetic elements to form BMRs when the tracking and pairing module operates using frame to frame comparisons.   The main problem resides on the fact that BMRs not only span multiple spatial scales, but also different temporal scales.  This difficult, but tractable exercise for a human supervisor becomes too complex for the current state of the art feature detection algorithms used in solar physics.  However, the existence of a human vetted BARD catalog lends itself perfectly to the role of anchor for a SWAMIS catalog of unpaired positive and negative magnetic elements.   Our plan for the creation of this hybrid catalog goes as follows:

\begin{enumerate}
  \item Create a SWAMIS catalog of tracked elements for each of the instruments used by the BARD catalog at full cadence.
  \item For each pair of subsequent magnetograms used by the BARD catalog identify the positive and negative elements, as well as their associated multi-scale components, that have been identified as linked.
  \item Using SWAMIS tracking information follow the evolution of the identified regions to see if there are mismatches connecting the two frames.
  \item Update SWAMIS flux element IDs in order to better match BARD constituencies, or flag the event for human supervision in case of failure.
  \item Create internal links marking the identified positive/negative pair in BARD and SWAMIS.
\end{enumerate}

\section{Summary and Concluding Remarks}\label{sec_conclusions}

In this paper we have reported the status of our effort to create a homogeneous BMR catalog by applying the BARD automatic detection code to magnetograms taken by the KPVT/512 (1976-1993), KPVT/SPMG (1992-1999), SOHO/MDI (1996-2011) and SDO/HMI (2010-2016) instruments.  In order to increase the accuracy of the catalog, we couple the automatic detection of BMRs with a human supervision module that allows a human observer to correct mislabels (introduced by errors in the automatic tracking module) and incorrect pairing of positive and negative regions to form BMRs (common when dealing with highly complex magnetic regions).  This arrangement allows us to take advantage of the strengths of automatic detection (consistency) and of manual detection (optimal detection of BMRs close enough to form active complexes).

The resulting catalog contains nearly 10,000 unique objects fully characterized and tracked at a cadence of one measurement per day, 10\% of which required some form of human intervention.  Our future plans involve the combination of the BARD and SWAMIS catalog to take advantage of the full cadence of the SOHO/MDI and SDO/HMI instruments, and the detection of magnetic features at smaller scales than those in the current BARD catalog.  The resulting composite catalog will be made public through the solar dynamo dataverse (https://dataverse.harvard.edu/dataverse/solardynamo) and will prove instrumental for enabling a new generation of statistical studies shedding light on the scaling and origin of surface magnetism, the nature of observable dynamo action, the dependence of surface magnetism on solar activity, and the characterization of the solar extended cycle.

\bibliographystyle{IEEEtran}
\bibliography{References}
%
%
%

\end{document}